\documentclass[trackchanges,twocolumn]{aastex701}

\usepackage{amsfonts,amsmath,amssymb}

\graphicspath{{./}{paper_figures/}}

\begin{document}

\title{High Dispersion Measure Fast Radio Bursts from Young, Dust-Poor H~II Regions}

\author[orcid=0009-0005-9389-9098,gname=Eliza, sname=Diggins]{Eliza C. Diggins}
\affiliation{Department of Astronomy, University of California Berkeley, Berkeley, CA, 94720, USA}
\email[show]{eliza.diggins@berkeley.edu}  

\author[orcid=0000-0002-4209-7408, gname=Calvin, sname=Leung]{Calvin Leung} 
\affiliation{Department of Astronomy, University of California Berkeley, Berkeley, CA, 94720, USA}
\affiliation{Department of Physics, University of California Berkeley, Berkeley, CA, 94720, USA}
\affiliation{Miller Institute of Basic Research, University of California Berkeley, Berkeley, CA, 94720, USA}
\email{calvin_leung@berkeley.edu}

\author[orcid=0000-0002-1568-7461, gname=Wenbin, sname=Lu]{Wenbin Lu}
\affiliation{Department of Astronomy, University of California Berkeley, Berkeley, CA, 94720, USA}
\affiliation{Theoretical Astrophysics Center, University of California Berkeley, Berkeley, CA, 94720, USA}
\email{wenbinlu@berkeley.edu}

\begin{abstract}
Several repeating fast radio burst (FRB) sources exhibit host-frame dispersion measure (DM) excesses far exceeding those expected from the diffuse ionized interstellar medium in their host galaxies, implying a substantial contribution from the $\lesssim$ kiloparsec-scale local environment. 
These excesses are often attributed to an expanding supernova remnant; however, this canonical interpretation does not explain their apparent preference for low-metallicity dwarf galaxy hosts, suggesting that the mechanism producing the local DM is coupled to the properties of the host galaxy.
In this work, we investigate whether the H~II regions generated by young, massive stellar associations can simultaneously explain these DM excesses and their preferential association with low-metallicity dwarf galaxies.
We develop a semi-analytic model for FRB-emitting neutron stars in cluster-forming molecular clouds and perform idealized one-dimensional radiation-hydrodynamic simulations using time-dependent ionizing luminosities from \texttt{pySTARBURST99}. 
Our findings indicate that high-mass $(M_\star \sim 10^5-10^6\;M_\odot)$, young ($\lesssim 10\;{\rm Myr}$) stellar associations in low-metallicity environments (LMC-like or lower) can produce ${\rm DM}_{\rm HII} \gtrsim 10^3\;{\rm pc\;cm^{-3}}$ for FRBs that form early ($\lesssim 5$ Myr) and remain embedded in the ionized gas.
The low-mass/metallicity preference arises from dust: absorption of the Lyman continuum suppresses the ionized column at higher metallicity, making extreme local DMs preferentially generated in dust-poor, low-metallicity environments.
In nearby host galaxies, this mechanism may be tested through NUV and ${\rm H}\alpha$ observations of the H~II region spatially coincident with the FRB.
Conversely, at cosmological distances, selection of a metal-rich, massive, and quiescent host galaxy sample where such extreme local contributions are likely to be suppressed may yield a cleaner FRB sample for using extragalactic DMs as tracers of the intervening intergalactic and circumgalactic media.
\end{abstract}



\section{Introduction}
\label{sec:Intro}
Fast radio bursts (FRBs) are millisecond-duration radio transients~\citep{2019A&ARv..27....4P}, widely believed to originate from magnetospheric processes associated with magnetars \citep[e.g.,][]{2020Natur.587...59B}, although their emission mechanism remains uncertain \citep[see, e.g.,][]{2019MNRAS.485.4091M, margalitFastRadioBursts2019, 2020MNRAS.498.1397L, zhang_physical_2020, beloborodovBlastWavesMagnetar2020, lyubarskyFastRadioBursts2020}. 
Despite their mysterious origin, FRBs have emerged as powerful probes of baryonic structure and feedback through measurements of their dispersion measures (DMs), which trace the integrated column density of free electrons along cosmological sightlines~\citep[e.g.,][]{2020Natur.581..391M}. In so doing, FRB DMs provide a unique window into the distribution of ionized gas in the intergalactic medium~\citep[e.g.,][]{2025NatAs...9.1226C,leung2025nulling}, circumgalactic medium~\citep[e.g.][]{leung2025stellar,mccarty2026cgm}, intervening sightlines~\citep{2023ApJ...954L...7L}, and sub-host scale local environments~\citep{piroDispersionRotationMeasure2018,orr_objects_2024}.

A central challenge in the use of FRBs as cosmological probes is disentangling the various contributions to the observed DM, including those from the Milky Way, the intergalactic medium, intervening structures, the host galaxy, and the immediate environment of the source \citep[e.g.,][]{cordesRedshiftEstimationConstraints2022, petroffFastRadioBursts2022}.
This task is complicated by the uncertain nature of FRB progenitors and their environments.
The existence of an FRB subpopulation with unusually large dispersion measures ($\sim 10^3\;{\rm pc\;cm^{-3}}$) at low redshift indicates that both small-scale and large-scale host-galaxy environments can contribute significantly to the total DM \citep{niuRepeatingFastRadio2022, bhandariNonrepeatingFastRadio2023, moroianuMilliarcsecondLocalizationAssociates2025}.
While the full FRB population does not cleanly trace host star formation rate, stellar mass~\citep{loudasUnveilingOriginFast2025, 2026ApJ...996...78H} or host galaxy integrated metallicity~\citep{2026ApJ...997L...6Y}, four of the five known ``hyperactive repeaters'' (FRB 20121102, 20190520B, 20190417A, and 20240114A) reside in low-metallicity dwarf galaxies --- a clue into the nature of this subpopulation.
These sources  stochastically emit frequent, narrow-band bursts ($\Delta \nu/\nu \sim 0.3$) with a broken power-law rate distribution~\citep{2024MNRAS.52710425S, kirstenLinkRepeatingNonrepeating2024}.
Though the sample of hyperactive repeaters remains small, the origin of their large source-frame dispersion measures ($\sim 100$--$1000\;{\rm pc\;cm^{-3}}$), as well as their apparent preference for low-mass hosts, remains poorly understood.~\citet{leung2025stellar} reported and~\citet{athukoralalage2026local} extended the $\mathrm{DM_{host}}-M^*$ relation: a decline in the host-frame DM with increasing host-galaxy stellar mass over a broad range of host stellar masses ($\log_{10}M^* = 8.5-11.4$). Coupled with the emerging association between dwarf galaxies and high-DM FRBs, this finding suggests a physical connection between the properties of the host galaxy and the origin of the dispersion measure excess. 

One possibility is that these sources represent a distinct population of young neutron stars formed preferentially in low-metallicity environments, analogous to the observed preference of long gamma-ray bursts and Type Ibc broad line supernovae for dwarf galaxies \citep{Bloom2002Observed, fruchterLongGrayBursts2006, 2008ApJ...673..999P}.
However, reproducing the large observed dispersion measures within the framework of an extreme parsec-scale environment, such as a magnetar wind nebula or supernova remnant around the central engine, requires a combination of extreme conditions, including massive progenitors ($M \gtrsim 10\,M_\odot$) with high mass loss rates, a dense ambient ISM ($n \gtrsim 10^2\;{\rm cm^{-3}}$), and very young source ages ($t \lesssim 10^3$ yr).
These conditions may be difficult to realize generically \citep[e.g][]{piroImpactSupernovaRemnant2016, zhaoFRB190520BEmbedded2021, 2017ApJ...841...14M, piroDispersionRotationMeasure2018} and may require a more extreme source class to be formed in dwarf galaxies.

An alternative scenario is that FRB progenitors do not ``know'' about their host galaxies, and as proposed by~\citep{athukoralalage2026local}, that the large dispersion measures instead arise from the local ($\lesssim 1\;{\rm kpc}$) environment of the source, for example, through ionization of the surrounding medium by massive stars or clustered star formation activity. 
Many FRBs are co-localized with star-forming regions in their host galaxies \citep[e.g.,][]{piroFastRadioBurst2021, 2022AJ....163...69B, dongMappingObscuredStar2024, tendulkar2017host}.
In this work, we explore this possibility with both analytical and numerical modeling of FRBs born in stellar associations generating large H~II regions.

The remainder of this paper is organized as follows. In
Section~\ref{sect:modeling}, we develop a semi-analytic model for the
dispersion-measure contribution from FRB progenitors embedded in natal H~II
regions. In Section~\ref{sect:numerical}, we describe idealized
one-dimensional radiation-hydrodynamic simulations that follow the response of
the ionized gas to the time-dependent ionizing luminosity of an evolving
stellar population. In Section~\ref{sect:observational_data}, we briefly describe a volume-limited set of FRBs with host-galaxy metallicity measurements which are compared with our models. In Section~\ref{sect:results}, we present the resulting
DM distributions across representative stellar population masses,
metallicities, and progenitor drift velocities, and compare
them with observed high-DM FRBs. In Section~\ref{sect:disc}, we discuss the
observational implications, uncertainties, and limitations of this scenario,
before summarizing our conclusions in Section~\ref{sect:conclusion}.

\begin{figure*}[ht]
    \resizebox{\hsize}{!}{
    \includegraphics[width=1\linewidth]{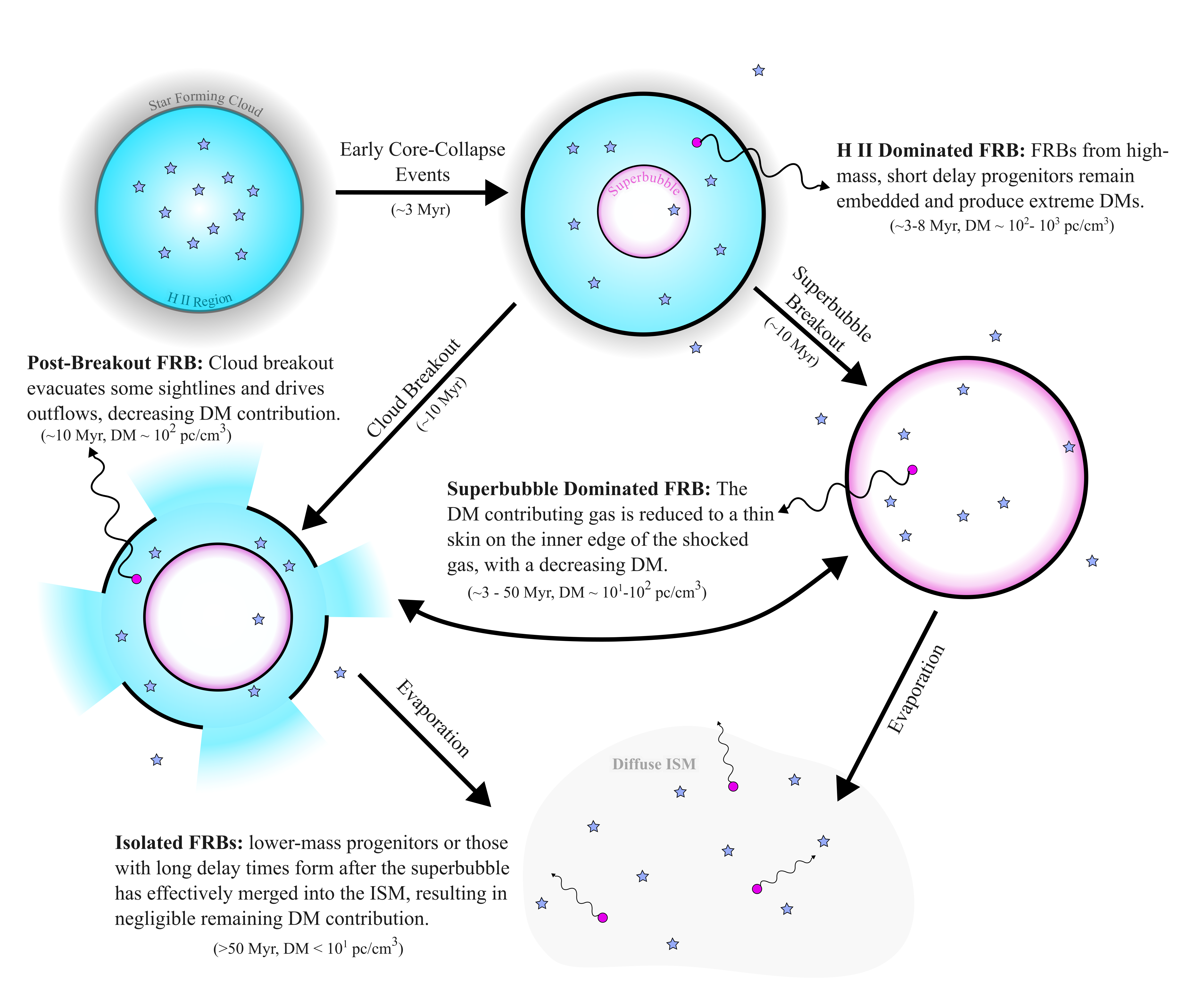}
   }
\caption{
Schematic evolution of a young massive stellar association embedded in a cluster-forming molecular cloud, illustrating the principal FRB environments considered in this work.
During the early H~II-dominated phase, neutron stars formed in the first core-collapse events remain embedded within the dense ionized gas and can acquire large association-scale dispersion measures.
The subsequent evolution can proceed through either photoionization-driven breakout of the H~II region from the parent cloud or the growth of a mechanically driven superbubble.
Cloud breakout opens low-density escape channels and reduces the ionized-gas column, while superbubble expansion evacuates the central volume and sweeps the gas into a dense shell whose DM contribution declines as the shell expands. 
At later times, after the natal cloud has dispersed or the source has dynamically separated from the association, the FRB enters an isolated-remnant phase in which the local dispersion measure is expected to be dominated by its supernova remnant, circumstellar environment, or the diffuse host-galaxy ISM.
The pathways shown are schematic and need not occur in a unique temporal sequence.
}
\label{fig:diagram}
\end{figure*}

\section{Modeling FRB Dispersion Measures in Natal H~II Regions}
\label{sect:modeling}

\begin{deluxetable*}{lccc}
\tabletypesize{\scriptsize}
\tablecaption{Fiducial parameters for the semi-analytic H~II-region model.
\label{tab:fiducial_analytic_parameters}}
\tablehead{
\colhead{Parameter} &
\colhead{Symbol} &
\colhead{Fiducial value} &
\colhead{Reference}
}
\startdata
GMC Stellar Mass&
$M_{\rm \star}$ &
$10^6\,M_\odot$ &
\citep{portegies2010young} 
\\
Star-formation efficiency &
$\eta$ &
$0.1$ &
\citep{2003ARAA..41...57L}
\\
Cloud Surface Density &
$\Sigma_{\rm cl}$ &
$10^2\;{\rm M_\odot\,pc^{-2}}$ &
\citep{larsonTurbulenceStarFormation1981}\\
Specific ionizing photon rate &
$Q_{\rm Lyc}$ &
$10^{46}\,{\rm s^{-1}}\,M_\odot^{-1}$ &
\citep{2025ApJS..280....5H}
\\
Ionized-gas temperature &
$T_{\rm II}$ &
$10^4\,{\rm K}$ &
\citep{draine2011physics}
\\
Dust absorption cross section &
$\langle\sigma_{\rm dust}\rangle_{\rm H}$ &
$2\times 10^{-21}\,{\rm cm^2}$ &
\citep{2001ApJ...548..296W}
\\
Plummer compactness factor &
$f_a$ &
$0.02$ &
\citep{ Brown_2021}
\\
Neutron-star mass range &
$[M_{\rm NS,min},M_{\rm NS,max}]$  &
$[9,60]\,M_\odot$ &
\citep{heger_how_2003, 2016ApJ...821...38S}
\\
Minimum FRB delay time &
$\tau_{\rm min}$ &
$10\,{\rm yr}$ &
\citep{2017ApJ...841...14M, 2019MNRAS.485.4091M}
\\
FRB activity timescale &
$\tau_{\rm frb}$ &
$10^4\,{\rm yr}$ &
\citep{2013BrJPh..43..356M, 2017ARAA..55..261K}
\\
Progenitor drift velocity &
$v_{\rm drift}$ &
$5\,{\rm km\,s^{-1}}$ &
\citep{portegies2010young, 2024MNRAS.533..705W}
\\
SNe Mechanical Efficiency & $\epsilon_{\rm mech}$ & $0.1$ & \citep{2014MNRAS.443.3463S,2017MNRAS.465.1720Y,2018MNRAS.481.3325F}\\
ISM Density & $n_{\rm H, \;ISM}$ & $1\;{\rm cm}^{-3}$ & \citep{draine2011physics}\\
Natal Kick Velocity & $v_{\rm kick}$ & $250\;{\rm km/s}$ & \citep{2005MNRAS.360..974H}\\
\enddata
\tablecomments{
The fiducial cloud quantities correspond to the high-mass GMC/cluster-forming
case adopted in the analytic model. The stellar mass is
$M_\star=\eta M_{\rm cl}$, and the total ionizing photon production rate is
$\mathbb{N}_{\rm Lyc}=Q_{\rm Lyc}M_\star$. The parameter
$f_a$ describes the compactness of the massive-star distribution relative to
the parent cloud; larger values correspond to more extended OB associations or
star-forming complexes. References should be interpreted as observational or
theoretical motivation for the adopted parameter scale rather than as unique
measurements of the fiducial values.
}
\end{deluxetable*}

To assess whether natal H~II regions can account for the large dispersion
measures observed in some FRBs, we construct a semi-analytic framework for
modeling young compact progenitors embedded in the ionized gas surrounding massive
stellar associations. 
We focus on FRBs produced by young neutron stars formed
in the core collapse of massive stars within cluster-forming molecular clouds.
In this picture, the relevant electron column is not set only by the
properties of the compact remnant or its immediate supernova environment, but
also by the larger-scale H~II region maintained by the surrounding population of
massive stars.
Figure~\ref{fig:diagram} illustrates the various dynamical considerations
that can influence the observed DM, including the time-dependent ionizing
photon budget, the drift of progenitors from their birth sites, and the
evolution of the H~II region over time.

The model consists of three coupled components:
\begin{enumerate}
    \item \textbf{FRB progenitor population synthesis.} 
    We connect the stellar population to the formation of FRB-capable compact remnants by sampling massive-star progenitors from the IMF, assigning core-collapse times through a mass--lifetime relation, and evolving their positions through progenitor drift, natal kicks, and post-collapse activation delays.

    \item \textbf{Environmental evolution.}
    We follow the expansion of the H~II region within its natal molecular cloud, its eventual breakout into the surrounding ISM, and the formation of a mechanically driven superbubble.

    \item \textbf{Dispersion-measure calculation.}
    At the time of FRB activation, we determine the source position and dynamical state of the H~II region, draw a randomized viewing direction, and integrate the free-electron column through the local environment to determine the DM.
\end{enumerate}

We combine these components in a Monte Carlo realization of the semi-analytic model.
For each synthetic source, we draw a progenitor mass and birth position, evolve the progenitor through core collapse and any subsequent activation delay, and determine its location relative to the evolving ionized gas. 
We then draw a viewing direction and integrate the free-electron density along the forward sightline through the model environment. Repeating this procedure over many realizations produces a distribution of local dispersion measures that captures variations in progenitor properties, source position, activation time, environmental evolution, and viewing geometry.

\subsection{The Natal Cloud and FRB Population Synthesis}

We begin by constructing a simple model for the temporal distribution of FRB progenitors within a coeval stellar population. In the case of star formation within molecular clouds, it is convenient to introduce the total cloud mass, $M_{\rm cl}$, and associated star-formation efficiency, $\eta$, such that the total stellar mass of the population is $M_{\rm \star} = \eta M_{\rm cl}$. Unless otherwise stated, we assume in this work that $\eta = 0.1$. 
This choice for $\eta$ represents the high end of plausible star forming efficiencies \citep{2003ARAA..41...57L} required to produce the most massive associations \citep{2016ApJ...822...52R}, with more typical clusters having $\eta \sim 0.01.$
Because we are primarily interested in unusually massive, efficiently cluster-forming environments, we adopt $\eta=0.1$ as a fiducial upper-end value rather than as representative of the GMC population as a whole. The predicted H~II-region dispersion measure depends only weakly on this choice (approximately ${\rm DM}\propto\eta^{1/6}$ in the analytic model; see section~\ref{subsect:HII}), so more typical efficiencies do not qualitatively alter our conclusions.

The stellar population is assumed to form in an instantaneous burst following a \citet{2001MNRAS.322..231K} initial mass function (IMF), $\xi(M)$, normalized such that $\xi(M)\,dM$ gives the number of stars formed in the mass interval $(M, M+dM)$ per unit stellar mass formed. To predict the expected number of FRBs present in the association, we adopt a fiducial zero-age main sequence (ZAMS) mass range for neutron star formation of $M \in [M_{\rm NS,min}, M_{\rm NS,max}] = [9,60]\,M_\odot$ \citep[e.g.,][]{2016ApJ...821...38S, heger_how_2003, 2008MNRAS.386L..23B}. The upper boundary is motivated by the existence of neutron stars with ZAMS masses $\gtrsim 50 M_{\odot}$~\citep[e.g.,][]{2006ApJ...636L..41M,2005ApJ...622L..49F,2009ApJ...707..844D}, but we note that the exact value is uncertain and may depend on metallicity, rotation, and explosion physics \citep[e.g.,][]{horvathBirthEventsMasses2022,woosleyBirthFunctionBlack2020, boccioliRemnantMasses1D2024}.
To connect stellar mass to main-sequence lifetime, we adopt the fits of
\citet{1989A&A...210..155M} (see also \citealt{2005A&A...430..491R}),
which are suitable for massive stars. Defining $m \equiv M/M_\odot$, the
main-sequence lifetime is
\begin{equation}
    \begin{aligned}
            \log_{10}&\left(\frac{\tau_{\rm MS}}{\rm yr}\right)
    = \\
    &\begin{cases}
        -0.6545 \log_{10} m + 10.00,
            & m < 1.3, \\[3pt]
        -3.70 \log_{10} m + 10.35,
            & 1.3 \leq m < 3, \\[3pt]
        -2.51 \log_{10} m + 9.77,
            & 3 \leq m < 7, \\[3pt]
        -1.78 \log_{10} m + 9.17,
            & 7 \leq m < 15, \\[3pt]
        -0.86 \log_{10} m + 8.06,
            & 15 \leq m < 60, \\[3pt]
        \log_{10}\left(1.2m^{-1.85} + 0.003\right) + 9,
            & m \geq 60.
    \end{cases}
    \end{aligned}
    \label{eq:stellar_lifetime}
\end{equation}

Using this mapping, we can express the neutron-star formation rate as a function of time. For a simple stellar population of total mass $M_{\rm \star}$, the formation rate is
\begin{equation}
\label{eq:ns_birth_rate}
\frac{dN_{\rm NS}}{dt}(t) = M_{\rm \star}\,B_{\rm NS}(t),
\end{equation}
where $B_{\rm NS}(t)$ is a \emph{neutron-star birth kernel} that encodes the temporal distribution of neutron-star-forming core-collapse events. This kernel is obtained by propagating the IMF through the mass–lifetime relation,
\begin{equation}
\label{eq:ns_kernel}
B_{\rm NS}(t) =
\begin{cases}
\xi(m)\,\left|\dfrac{dm}{dt}\right|, & m(t)\in[M_{\rm NS,min},M_{\rm NS,max}],\\[8pt]
0, & \text{otherwise.}
\end{cases}
\end{equation}

To map neutron-star formation onto the temporal distribution of FRB events, we introduce an effective delay / activation kernel.
We define $\Psi(\Delta t)$ as the \emph{FRB event kernel}, which specifies the relative burst rate at a time $\Delta t$ after core collapse.
The rate of FRB events in the population at a time $t$ is then 
\begin{equation}
\frac{dN_{\rm FRB}}{dt}
\propto M_{\rm \star}
\int_0^t
B_{\rm NS}(\tau)
\Psi(t-\tau)\,d\tau,
\end{equation}
where the constant of proportionality encodes the unknown fraction of neutron stars which eventually produce FRBs.
This formulation is intentionally agnostic to progenitor physics and emission mechanism, with different progenitor scenarios corresponding to different choices of $\Psi(\Delta t)$.

In this work, we adopt a simple scenario in which FRB activation occurs promptly following core collapse, which is a common feature of many FRB models \citep[e.g.,][]{2019MNRAS.485.4091M,2017ApJ...841...14M,2017ApJ...842...34W}.
In this picture, FRB emission is possible immediately after core collapse, with the probability of activity declining smoothly with source age.

We model this behavior using an exponential activity kernel,
\begin{equation}
\label{eq:psi_prompt}
\Psi_{\rm prompt}(\Delta t) \propto
\Theta(\Delta t - \tau_{\rm min})\exp\!\left(-\frac{\Delta t}{\tau_{\rm frb}}\right),
\end{equation}
where $\tau_{\rm frb}$ is an effective activity timescale, $\tau_{\rm min}$ is the minimum delay time, and $\Theta$ is the Heaviside step function.
The latter accounts for suppression of observable FRB emission at very early times due to free--free absorption in the immediate post-supernova environment, and is fixed to $\tau_{\rm min}=10\,{\rm yr}$ following \citet{2017ApJ...841...14M}.

We adopt a fiducial value $\tau_{\rm frb}=10^{4}\,{\rm yr}$, motivated by models of magnetar evolution in which the magnetic energy available to power flaring activity persists for $\sim10^{4}$--$10^{5}\,{\rm yr}$ after neutron-star formation \citep[e.g.,][]{2013BrJPh..43..356M,2017ARAA..55..261K}.
We note however that population arguments for the globular-cluster source FRB~20200120E have suggested a longer repeating-activity lifetime, $10^{4}\,{\rm yr}\lesssim\tau_{\rm frb}\lesssim10^{6}\,{\rm yr}$, with a preferred value of order $10^{5}\,{\rm yr}$ \citep{luImplicationsRapidlyVarying2021}.
Because the H~II-region radius and density evolve primarily on Myr timescales, varying $\tau_{\rm frb}$ within this range has only a modest effect on the predicted DM distribution for the prompt-formation channel considered here (see Section~\ref{sect:mass_and_dynamics} for in-depth discussion).

\subsection{Stars: Dynamical Dispersal of FRB Progenitors}
Prior to core collapse, FRB progenitors need not remain fixed at the centers of their natal H~II regions. Massive stars form in spatially extended and often substructured associations, and their subsequent motions gradually reduce the spatial correlation between the progenitor and its star forming environment.
To model the spatial distribution of FRB progenitors in our model, we approximate the natal stellar population as a Plummer sphere with density profile
\begin{equation}
\rho_\star(r) =
\frac{3M_{\rm \star}}{4\pi a^3}
\left(1+\frac{r^2}{a^2}\right)^{-5/2},
\end{equation}
where $M_{\rm \star}$ is the population mass and $a$ is the Plummer scale radius \citep{1911MNRAS..71..460P}. Rather than treating the Plummer scale radius as an independent dimensional parameter, we tie the size of the stellar association to the parent molecular cloud. 

For star-forming molecular clouds, we adopt a fiducial surface density $\Sigma_{\rm cl}=10^2\;{\rm M_\odot\,pc^{-2}}$, characteristic of GMCs in the inner Milky Way \citep{heyerReExaminingLarsonsScaling2009,rosolowskyGiantMolecularCloud2021}. Early studies of Galactic molecular clouds suggested that $\Sigma_{\rm cl}\sim10^2\;{\rm M_\odot\,pc^{-2}}$ was approximately universal \citep{larsonTurbulenceStarFormation1981}; however, subsequent Galactic and extragalactic surveys have demonstrated substantial environmental variation \citep{heyerReExaminingLarsonsScaling2009,rosolowskyGiantMolecularCloud2021,2018ApJ...860..172S}, with characteristic cloud-scale surface densities spanning roughly $10$--$200\;{\rm M_\odot\,pc^{-2}}$ in nearby disk galaxies.
Assuming a uniform spherical geometry (using $\mu_{\rm H} = 1.4$ for the mean molecular weight per hydrogen atom)
\begin{equation}
    \label{eq:n_H_scale}
    \begin{aligned}
        n_{H,0}& \sim \frac{3}{4\mu_{ H} m_H} \pi^{1/2} \Sigma_{\rm cl}^{3/2}\eta^{1/2} M_{\rm \star}^{-1/2}\\
               & \sim 12 \;\eta_{-1}^{1/2} M_{\star,\rm 6}^{-1/2}\Sigma_{\rm cl,2}^{3/2}\;{\rm cm^{-3}}  
    \end{aligned}
\end{equation}
and a corresponding cloud radius
\begin{equation}
    \label{eq:R_cl_scaling}
    R_{\rm cl} = \sqrt{\frac{M_{\rm cl}}{\pi \Sigma_{\rm cl}}} \approx 178\;{\rm pc} \;\eta_{-1}^{-1/2}M_{\rm \star,6}^{1/2} \Sigma_{\rm cl,2}^{-1/2}
\end{equation}
Because star formation occurs deeply embedded within the parent cloud, we parameterize the Plummer scale radius $a$ such that $a = f_a R_{\rm cl}$, where $f_a$ is fiducially taken to be 0.02, producing typical clusters with core radii $\sim 1-5\;{\rm pc}$ \citep{Brown_2021}. $f_a$ may vary significantly depending on context, with looser OB associations being much more extended ($f_a\sim 0.1$) \citep{portegies2010young, Wright2020OBAAA}.

We emphasize that the Plummer profile is adopted as a phenomenological spatial kernel rather than as a dynamical model of the association. Star forming clouds are often substructured, anisotropic, and dynamically young, with kinematics that may reflect hierarchical star formation in the natal molecular cloud rather than relaxation in a bound stellar potential \citep[e.g.,][]{2020MNRAS.495..663W, 2018MNRAS.475.5659W, Lim2019AGVF,2024MNRAS.533..705W}. Nevertheless, we adopt a Plummer profile as a minimal analytic prescription that captures the critical features of the stellar association without introducing additional parameters that would not be well constrained by current observations. We therefore interpret $a$ as an effective size parameter for the progenitor distribution, not as the scale radius of a relaxed, virialized stellar system. 

FRB progenitors may also drift relative to their star-forming environment over their main-sequence lifetimes. We describe this effect using an effective drift velocity, $v_{\rm drift}$, which parameterizes the loss of spatial correlation between the progenitor and its birth site prior to core collapse. It is taken to represent the coarse-grained relative motion expected in young associations due to inherited turbulent velocities, association-scale expansion, and dispersal following gas removal. Motivated by observed velocity dispersions and expansion speeds in young stellar clusters and OB associations, we consider fiducial values $v_{\rm drift}=1$--$10\;{\rm km\,s^{-1}}$ \citep[see e.g.][]{portegies2010young,2024MNRAS.533..705W, 2018MNRAS.480..800H}. The lower end of this range corresponds to modest internal motions in loosely bound or unbound associations, while the upper end should be regarded as an aggressive dispersal case rather than a typical value. Our fiducial value is $v_{\rm drift} = 5\;{\rm km/s}$, assigned to each progenitor with random orientation relative to the natal association.

For neutron star progenitors with masses between $9\;{\rm M}_\odot$ and $60\;{\rm M}_\odot$, the main sequence lifespans range from $3.6\;{\rm Myr}$ to $29.6\;{\rm Myr}$, and the fiducial drift displacement during the main sequence is bounded,
\begin{equation}
    18.4 \left(\frac{v_{\rm drift}}{1\;{\rm km/s}}\right)\;{\rm pc} \le \Delta r_{\rm MS} \le 151.3\left(\frac{v_{\rm drift}}{1\;{\rm km/s}}\right){\rm pc}.
\end{equation}
As a fraction of the cloud radius,
\begin{equation}
    \label{eq:drift_to_cloud_ratio}
    \left[0.1 \leq \frac{\Delta r_{\rm MS}}{R_{\rm cl}}\le 0.85 \right] \eta_{-1}^{1/2} M_{\star,6}^{-1/2}\Sigma_{\rm cl,2}^{1/2}  \left(\frac{v_{\rm drift}}{5\;{\rm km/s}}\right),
\end{equation}
where coefficients are to be applied to either side of the bracket.
Equation~\eqref{eq:drift_to_cloud_ratio} demonstrates that low-mass neutron star progenitors $(M\sim 9\;{\rm M}_\odot)$ may be dissociated from their birth region quite easily during their main-sequence lifetime, requiring velocities of $\sim 5\;{\rm km/s}$ to do so. The most massive progenitors, by contrast, collapse before drifting far from their birth sites and are therefore preferentially selected into the high-DM population.

In addition to the displacement accumulated during the progenitor's main-sequence lifetime, neutron stars receive natal kicks at core collapse.
Observed pulsar velocity distributions imply characteristic kick speeds of order $\sim250\;{\rm km\,s^{-1}}$ \citep{2005MNRAS.360..974H,2025NewAR.10101734P}.
We therefore include natal kicks explicitly in the Monte Carlo model by drawing an isotropic kick velocity for each newly formed neutron star and propagating it for the interval between core collapse and FRB activation.

For an activation delay $\Delta t$, the resulting displacement is
\begin{equation}
    \Delta r_{\rm kick}
    \simeq
    2.6\;{\rm pc}
    \left(\frac{v_{\rm kick}}{250\;{\rm km\,s^{-1}}}\right)
    \left(\frac{\Delta t}{10^4\;{\rm yr}}\right).
\end{equation}
Thus, for the prompt-activation scenario adopted in our fiducial model,
$\Delta t\lesssim10^2$--$10^4\,{\rm yr}$, natal kicks generally produce only
parsec-scale offsets and remain subdominant to the displacement accumulated
before core collapse.
It should, however, be noted that natal kicks have also been invoked to explain FRBs observed at substantial
offsets from nearby star-forming regions
\citep[e.g.,][]{2021ApJ...908L..12T,2025ApJ...989L..48C}. Such systems
therefore require longer post-collapse delay times than those emphasized in
our prompt-activation model. Binary evolution may provide an additional source
of displacement through pre-collapse systemic recoil following the explosion
of a companion, but we do not model this effect here.

\subsection{Dynamics of an Idealized H~II Region}
\label{subsect:HII}

Having established the temporal and spatial distribution of FRB progenitors within their natal stellar population, we now model the ionized gas environment that sets the resulting dispersion measure.
The H~II region surrounding a newly formed stellar population is established rapidly following star formation, on approximately the recombination timescale. For typical conditions \citep[e.g.,][]{1997MNRAS.292...27H}, this is
\begin{equation}
    \begin{aligned}
    t_{\rm rec} &= \frac{1}{\alpha_B(T_{\rm II}) n_e} \\
    &\approx 1.4\times10^{4}\;{\rm yr}
    \left(\frac{T_{\rm II}}{10^4\;{\rm K}}\right)^{0.7}
    \left(\frac{n_e}{10\;{\rm cm}^{-3}}\right)^{-1},
    \end{aligned}
\end{equation}
where $\alpha_B$ is the case-B recombination coefficient, $T_{\rm II}$ is the temperature inside the H~II region, and $n_e$ is the electron number density. 
This timescale is much shorter than the lifetimes of massive stars, so the ionized region is effectively established well before the first FRB progenitors undergo core collapse.

\begin{figure}
    \includegraphics[width=\linewidth]{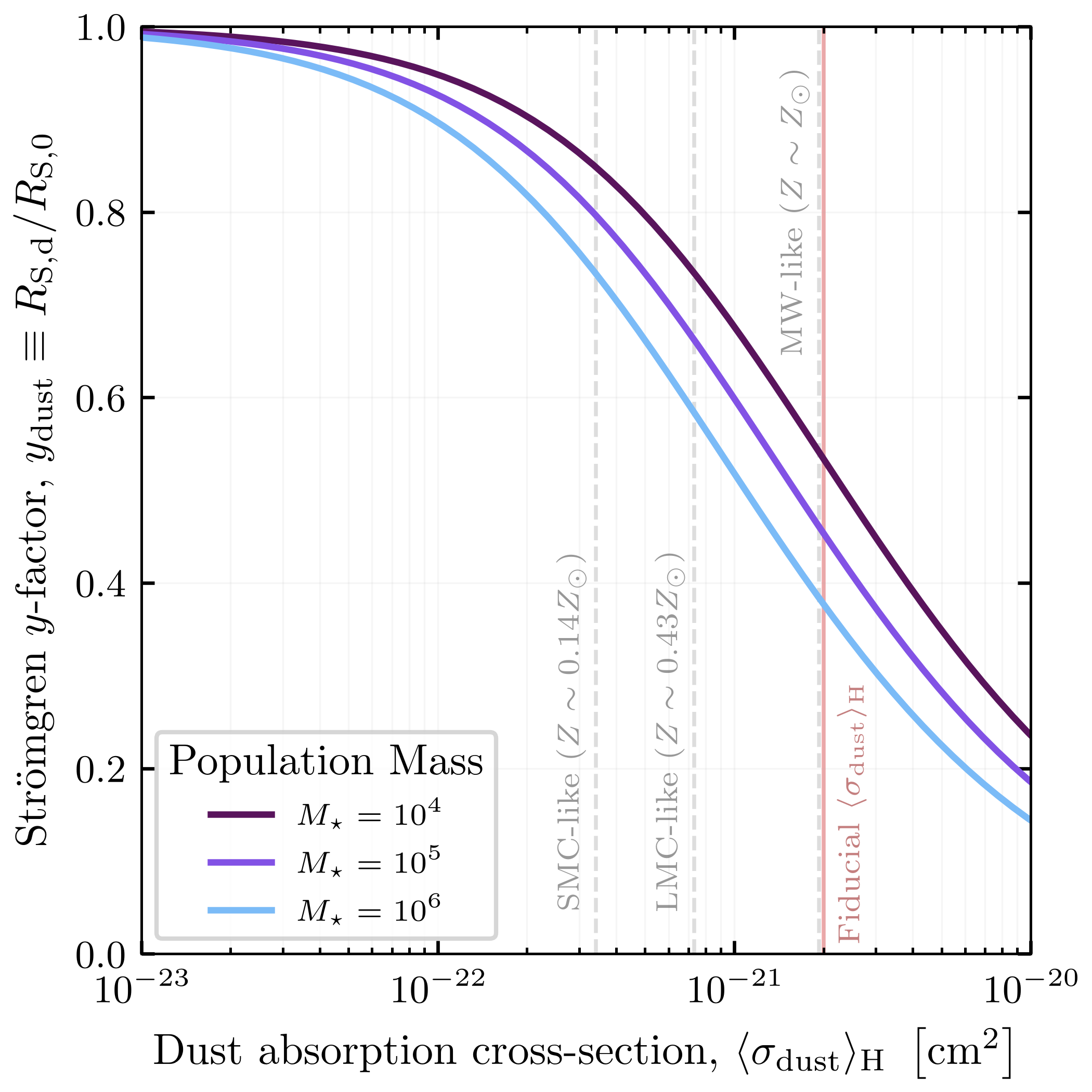}
    \caption{
    Str\"omgren $y$-factor (equation~\ref{eq:dusty_radius_suppression}) as a function of the dust absorption cross section to photoionizing radiation for three characteristic population masses, $M_\star$. Vertical lines indicate the photoionizing cross sections computed for the LMC, SMC, and Milky-Way from {\tt pySTARBURST99} and the grain models of \citet{2001ApJ...548..296W} using the method described in section~\ref{sect:numerical}. The red line indicates our fiducial value of $\langle\sigma_{\rm dust}\rangle_{\rm H} = 2\times 10^{-21}\;{\rm cm^2}$, roughly corresponding to the Milky-Way case.
    }
    \label{fig:stromgren_y_factor}
\end{figure}

Because the recombination timescale is much shorter than the dynamical
evolution timescale, the ionization state of the gas adjusts rapidly,
justifying the assumption of ionization equilibrium.
Under this approximation, the structure of the H~II region is set by a balance between photoionization, recombination, and dust absorption.
Let $\mathbb{N}_{\rm Lyc}$ denote the production rate of hydrogen-ionizing photons from the stellar population at the center of the cluster such that $N_{\rm Lyc}(r=0) = \mathbb{N}_\mathrm{Lyc}$, and let $N_{\rm Lyc}(r)$ be the ionizing photon rate that remains available at radius $r$. 
In our semi-analytic model, we assume $\mathbb{N}_{\rm Lyc}$ is \emph{constant in time}, but in the numerical simulations described in Section~\ref{sect:numerical}, we allow it to vary with the evolving stellar population.
By enforcing ionization equilibrium, the ionizing flux must obey
\begin{equation}
    \frac{dN_{\rm Lyc}}{dr}
    =
    -\underbrace{
    4\pi r^2 \alpha_B(T_{\rm II}) n_{\rm H}^2
    }_{\text{case-B recombinations}}
    -
    \underbrace{
    \left<\sigma_{\rm dust}\right>_{\rm H}
    n_{\rm H}
    N_{\rm Lyc}
    }_{\text{dust absorption}} .
    \label{eq:dusty_photon_budget}
\end{equation}
Here $n_{\rm H}$ is the hydrogen number density, and
$\langle\sigma_{\rm dust}\rangle_{\rm H}$ is the photon-number-weighted dust
absorption cross section per hydrogen nucleus for Lyman-continuum photons. In fiducial models, we adopt $\langle\sigma_{\rm dust}\rangle_{\rm H} \sim 2\times10^{-21}\;{\rm cm}^2$, corresponding to Milky-Way-like media \citep{2001ApJ...548..296W}.
It should be noted that this cross section is the absorption cross section only, as scattering of EUV photons does not eliminate them from the overall budget of ionizing radiation. 
In writing the recombination term as $\alpha_B n_{\rm H}^2$, we have assumed a
sharp ionization front and a nearly fully ionized interior, such that
$n_e \simeq n_p \simeq n_{\rm H}$. A partially ionized treatment would instead
replace this factor by $n_e n_p = x^2 n_{\rm H}^2$ for pure hydrogen. The
boundary of the H~II region occurs where the available ionizing photon budget is
exhausted,
\begin{equation}
    N_{\rm Lyc}(R_{\rm S,d}) = 0 .
    \label{eq:dusty_boundary_condition}
\end{equation}
Following the classical treatment of dusty Str\"omgren spheres \citep{1972ApJ...177L..69P}, for a uniform density interior, equation~\eqref{eq:dusty_photon_budget} may be solved implicitly in the form
\begin{equation}
    \mathbb{N}_{\rm Lyc} = \frac{4\pi \alpha_B(T_{\rm II})}{\langle \sigma_{\rm dust}\rangle_{\rm H}^3 n_{\rm H}} \left[e^{\tau_d}(\tau_d^2 - 2\tau_d + 2) -2\right],
    \label{eq:dust_stromgren_solution}
\end{equation}
where $\tau_d = \langle \sigma_{\rm dust}\rangle_{\rm H} n_H R_{\rm S,d}$ and $R_{\rm S,d}$ is the boundary of the ionized region.
In the dust-free case, the solution to this equation corresponds to the Str\"omgren radius,
\begin{equation}
R_{\rm S,0}
=
\left(
\frac{3\mathbb{N}_{\rm Lyc}}
{4\pi \alpha_B(T_{\rm II}) n_{\rm H}^2}
\right)^{1/3}.
\label{eq:dust_free_stromgren_radius}
\end{equation}
Using the cloud scalings of Equations~\eqref{eq:n_H_scale}--\eqref{eq:R_cl_scaling} and a specific ionizing flux $Q_{\rm Lyc} =\mathbb{N}_{\rm Lyc}/M_{\rm \star}$, the Str\"omgren radius may be written in the form
\begin{equation}
\label{eq:stromgren_sphere_scaling}
R_{S,0} \approx
128\;{\rm pc}\,
\eta_{-1}^{-1/3}
Q_{{\rm Lyc},46}^{1/3}
M_{\rm \star,6}^{2/3}
\Sigma_{\rm cl,2}^{-1}
T_{{\rm II},4}^{0.23}.
\end{equation}
Since $R_{\rm cl}=(M_{\rm cl}/\Sigma_{\rm cl}\pi)^{1/2}$,
\begin{equation}
\label{eq:stromgren_sphere_ratio_scaling}
    \frac{R_{S,0}}{R_{\rm cl}} = 0.72\;\eta_{-1}^{1/6}
Q_{{\rm Lyc},46}^{1/3}
M_{\rm \star,6}^{1/6}
\Sigma_{\rm cl,2}^{-1/2}
T_{{\rm II},4}^{0.23}.
\end{equation}
It is useful to express the dusty Str\"omgren radius in terms of the
dust-free value by defining
\begin{equation}
    y_{\rm dust}
    \equiv
    \frac{R_{\rm S,d}}{R_{\rm S,0}},
    \qquad
    0 < y_{\rm dust} \leq 1.
    \label{eq:dusty_radius_suppression}
\end{equation}
The corresponding dust optical depth across the ionized region is
\begin{equation}
    \tau_{\rm d}
    =
    \langle \sigma_{\rm dust}\rangle_{\rm H}
    n_{\rm H}
    R_{\rm S,d}
    =
    y_{\rm dust}\tau_{\rm S,0},
    \label{eq:dusty_region_tau}
\end{equation}
where
\[
\tau_{\rm S,0}
=
\langle \sigma_{\rm dust}\rangle_{\rm H}
n_{\rm H}
R_{\rm S,0}.
\]
Using the formal solution above together with the boundary condition
in Equation~\eqref{eq:dusty_boundary_condition}, $y_{\rm dust}$ is determined
implicitly by
\begin{equation}
    e^{\tau_{\rm d}}
    \left(
    \tau_{\rm d}^2 - 2\tau_{\rm d} + 2
    \right)
    -2
    =
    \frac{\tau_{\rm S,0}^3}{3}.
    \label{eq:dusty_stromgren_condition}
\end{equation} 
Figure~\ref{fig:stromgren_y_factor} shows the resulting suppression factor for our fiducial cloud and stellar-population parameters.

The ionized gas is overpressured relative to the surrounding neutral medium, driving expansion of the H~II region. Approximating the region as a uniform interior bounded by a thin swept-up shell, the shell evolution is governed by momentum conservation \citep{2002ApJ...566..302M},
\begin{equation}
\frac{d}{dt}\left(M_{\rm sh}\dot{R}_{\rm II}\right)
=
4\pi R_{\rm II}^2 P_{\rm II},
\label{eq:shell_momentum}
\end{equation}
where we neglect the external thermal pressure ($P_{\rm I} \ll P_{\rm II}$) on the basis that it is cool relative to the internal gas. The interior pressure is
\begin{equation}
\label{eq:HII_pressure}
P_{\rm II} = \frac{\mu_H}{\mu_{\rm II}} n_H k_B T_{\rm II},
\end{equation}
where $\mu_{\rm II}$ is the mean molecular weight of the ionized gas and we have neglected the contributions of radiation pressure (see \citealt{2011ApJ...732..100D} for discussion of radiation pressure in dusty H~II regions).
The swept-up shell mass is 
\begin{equation}
M_{\rm sh}(R_{\rm II}) =
4\pi \int_0^{R_{\rm II}} r^2 \rho_{\rm cl}(r)\,dr,
\label{eq:swept_mass}
\end{equation}
where $\rho_{\rm cl}$ is the density of the ambient cloud.
In the dust-free problem with constant ionizing luminosity and a uniform interior density, this evolution is described by the classical Spitzer solution \citep{1978ppim.book.....S,1954BAN....12..187K},
\begin{equation}
R_{\rm II}(t)
\simeq
R_{\rm S,0}
\left(
1+\frac{7c_s t}{4R_{\rm S,0}}
\right)^{4/7},
\label{eq:spitzer_solution}
\end{equation}
where $c_s=\sqrt{k_B T_{\rm II}/m_p \mu_{\rm II}}$ is the ionized-gas sound speed.

For later analytic estimates, it is useful to define a \emph{modified Spitzer
approximation} by replacing the dust-free equilibrium radius with the dusty
Str\"omgren radius,
\begin{equation}
    R_{\rm II,Sp,d}(t)
    \equiv
    R_{\rm S,d}
    \left(
        1+\frac{7c_s t}{4R_{\rm S,d}}
    \right)^{4/7}.
    \label{eq:modified_spitzer_solution}
\end{equation}
Unlike Equation~\eqref{eq:spitzer_solution}, however,
Equation~\eqref{eq:modified_spitzer_solution} is not an exact solution to
the dusty expansion problem, because dust absorption modifies the
ionization-balance relation that determines the interior density and
pressure throughout the evolution.
We therefore evolve the H~II region numerically using a thin-shell model in which the ionization front and the forward shock are assumed to remain coincident,
\begin{equation}
    R_{\rm IF} = R_{\rm sh} \equiv R_{\rm II}.
\end{equation}
The shell mass, radial momentum, and radius are evolved according to 
\begin{equation}
    \begin{aligned}
        \frac{dM_{\rm sh}}{dt}
        &=
        4\pi R_{\rm II}^{2}
        \rho_{\rm cl}(R_{\rm II})
        \dot{R}_{\rm II},
        \\
        \frac{dp_{\rm sh}}{dt}
        &=
        4\pi R_{\rm II}^{2}
        \left(P_{\rm II}-P_{\rm I}\right),
        \\
        \frac{dR_{\rm II}}{dt}
        &=
        \frac{p_{\rm sh}}{M_{\rm sh}},
    \end{aligned}
    \label{eq:dusty_shell_evolution}
\end{equation}
where $p_{\rm sh}=M_{\rm sh}\dot{R}_{\rm II}$. As above, we neglect the
external pressure in the fiducial calculations because
$P_{\rm I}\ll P_{\rm II}$.

The dynamical system is closed by determining the uniform ionized-gas density,
$n_{\rm H}$, at each timestep from the instantaneous dusty photon budget.
Specifically, for the current shell radius $R_{\rm II}$, we solve
Equation~\eqref{eq:dust_stromgren_solution} for $n_{\rm H}$ subject to
\begin{equation}
    N_{\rm Lyc}(R_{\rm II}(t))=0 \implies n_{\rm H, II}(t).
\end{equation}
The resulting density determines the interior pressure through equation~\eqref{eq:HII_pressure}.
Thus, each evaluation of the dynamical equations consists of first solving the
implicit ionization-balance condition for $n_{\rm H}$ and then using the
corresponding pressure to update the shell momentum, swept-up mass, and radius.
In the limit
$\langle \sigma_{\rm dust}\rangle_{\rm H}\rightarrow0$, this closure reduces to
\begin{equation}
    n_{\rm H}
    =
    \left(
    \frac{3\mathbb{N}_{\rm Lyc}}
    {4\pi\alpha_B(T_{\rm II})R_{\rm II}^{3}}
    \right)^{1/2},
\end{equation}
and the model recovers the usual dust-free thin-shell evolution.

\begin{figure}
    \centering
    \includegraphics[width=\linewidth]{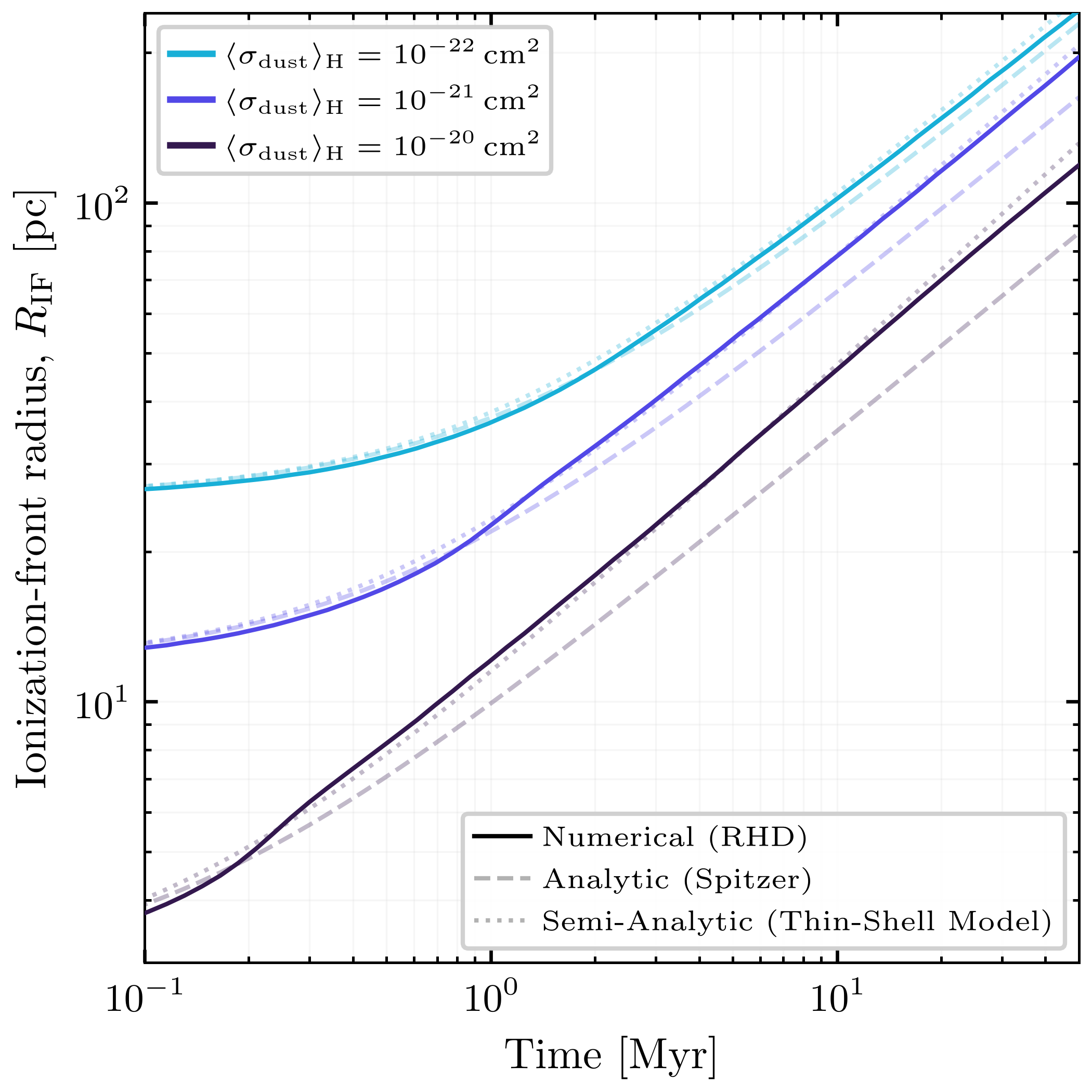}
    \caption{
    Solutions for the expansion of a fiducial H~II region with constant ionizing flux using the three techniques described in this work (RHD, semi-analytic, and analytic). Solutions were obtained using fiducial parameters and a population mass $M_\star = 10^6\;{\rm M_\odot}$ for three dust opacities (colored). In each case, the solution was calculated using the radiation hydrodynamics (RHD) method described in section~\ref{sect:numerical}, the semi-analytic method (see equation~\ref{eq:dusty_shell_evolution}), and the analytic, modified Spitzer solution (equation~\ref{eq:modified_spitzer_solution}). The numerical and semi-analytic solutions are each accurate with one another to $\lesssim 5\%$, while at high $\langle\sigma_{\rm dust}\rangle_{\rm H}$, the Spitzer solution can vary by considerably more from the RHD and semi-analytic solutions.
    }
    \label{fig:radhd_comparison}
\end{figure}

While we use this numerical scheme in our semi-analytic model, the Spitzer solution, adapted to use the dusty equilibrium radius (equation~\ref{eq:modified_spitzer_solution}), provides a useful (analytic) approximation for the expansion of the dusty H~II region and, for
our fiducial parameters, is accurate to within 10\% as late as $t \sim 10^7\;{\rm yr}$ for Milky-Way-like dust opacities. 
Figure~\ref{fig:radhd_comparison} illustrates the relative accuracy of our thin-shell model and the approximate Spitzer solution, along with the RadHD simulation solutions described below in section~\ref{sect:numerical}.
We therefore use the modified Spitzer solution as a convenient analytic approximation for deriving scaling relations and interpreting the evolution, with the understanding that it slightly underestimates the true ionized radius.

\subsection{The Dispersion Measure}

Within the H~II region, hydrogen is nearly fully ionized and dominates the dispersion measure of an embedded FRB. For a source located at position $\mathbf{r}$ relative to the center of the region, the path length $\ell > 0$ through the ionized gas along a line of sight $\hat{\mathbf{n}}$ is obtained from
\[
|\mathbf{r} + \ell \hat{\mathbf{n}}|^2 = R_{\rm II}^2,
\]
giving
\begin{equation}
\ell = -\mathbf{r}\cdot \hat{\mathbf{n}} + \sqrt{(\mathbf{r}\cdot \hat{\mathbf{n}})^2 + R_{\rm II}^2 - r^2}.
\end{equation}

Assuming a uniform density interior, the corresponding electron density may be determined from equation~\eqref{eq:dust_stromgren_solution}. For our analytic estimates, this is approximately 
\begin{equation}
    \label{eq:ne_approx_stromgren}
    n_e \approx n_{\rm H,0}\left(\frac{R_{\rm S,d}}{R_{\rm II}(t)}\right)^{3/2} = \left(\frac{3 y_{\rm dust}^3 \mathbb{N}_{\rm Lyc}}{4\pi \alpha_B R_{\rm II}^3}\right)^{1/2},
\end{equation}
where we have used the approximate form of equation~\eqref{eq:modified_spitzer_solution}.
The dispersion measure is
\[
{\rm DM} = \ell n_e.
\]
For a source located near the center of the region, the line-of-sight has length $R_{\rm II}(t)$ and therefore
\begin{equation}
\label{eq:dm_scaling}
\begin{aligned}
    {\rm DM}(t) \approx&
1.5\times10^{3}\;{\rm pc\;cm^{-3}}\,
\left(1+\frac{7c_s t}{4R_{\rm S,d}}\right)^{-2/7} \\&y_{\rm dust} \eta_{-1}^{1/6}Q^{1/3}_{\rm Lyc,46} M_{\rm \star,6}^{1/6} \Sigma_{\rm cl,2}^{1/2} T_{\rm II,4}^{0.23}.
\end{aligned}
\end{equation}
Thus, as the H~II region expands, the dispersion measure decreases despite the increasing path length, due to the declining electron density.

\subsection{Cloud Breakout}
\label{sect:cloud_breakout}

Immediately following star formation, the H~II region may already encompass an appreciable fraction of the natal cloud radius (Equation~\ref{eq:stromgren_sphere_ratio_scaling}). 
As it expands, it will eventually reach the edge of the cloud ($R_{\rm II}=R_{\rm cl}$) at a time approximately given by
\begin{equation}
    \begin{aligned}
        t_{\rm breakout}
        &\approx
        \frac{4R_{\rm S,d}}{7c_s}
        \left[
        \left(\frac{R_{\rm cl}}{R_{\rm S,d}}\right)^{7/4}
        -1
        \right],
        \\
        &\sim
        10^7\;{\rm yr}\;
        y_{\rm dust}^{-3/4}
        \eta_{-1}^{-5/8}
        Q_{\rm Lyc,46}^{-1/4}
        M_{\star,6}^{3/8}
        \Sigma_{\rm cl,2}^{-1/8}
        T_{\rm II,4}^{-0.67},
    \end{aligned}
    \label{eq:hii_breakout_time}
\end{equation}
for the fiducial cloud parameters.
Once the ionization front reaches the edge of the parent cloud, the subsequent evolution is no longer well described as that of a closed, spherical H~II region.
In a realistic molecular cloud, breakout is expected to occur first along low-density channels, allowing ionized gas and Lyman-continuum photons to escape and producing blister-like or champagne-flow morphologies \citep{1979A&A....71...59T,1981A&A....98...85B,1983A&A...127..313Y, 2007ASSP....1..103H,2019MNRAS.487.2200Z, 2025arXiv251102061H}.
The resulting electron column is therefore strongly geometry dependent: sightlines passing through evacuated channels may have substantially reduced dispersion measures, whereas those intersecting dense ionized structures or swept-up shells may retain large columns.

To incorporate the leading-order dynamical effect of cloud breakout into the semi-analytic model, we allow the ambient density entering Equation~\eqref{eq:dusty_shell_evolution} to change discontinuously at the nominal cloud boundary,
\begin{equation}
    \rho_{\rm init}(r)
    =
    \begin{cases}
        \rho_{\rm cl,0},
        & r \leq R_{\rm cl},\\[4pt]
        \rho_{\rm ISM},
        & r > R_{\rm cl},
    \end{cases}
    \label{eq:cloud_ism_density_profile}
\end{equation}
where $\rho_{\rm cl,0}=\mu_{\rm H}m_{\rm H}n_{\rm H,0}$ is the uniform cloud density (equation~\ref{eq:n_H_scale}) and $\rho_{\rm ISM}=\mu_{\rm H}m_{\rm H}n_{\rm H,ISM}$ is the density of the surrounding diffuse interstellar medium.
The swept-up shell mass and momentum are evolved continuously across $R_{\rm cl}$ using Equation~\eqref{eq:dusty_shell_evolution}. 
Thus, upon leaving the cloud, the shell retains the dense material accumulated during its earlier expansion, while its subsequent mass-loading rate decreases as it propagates into the lower-density external medium.
This will cause the expansion of the H~II region to accelerate and rapidly decrease the dispersion-producing electron column.

The sharp density transition in Equation~\eqref{eq:cloud_ism_density_profile}
should be regarded as an idealized representation of the cloud boundary.
Observed and simulated molecular clouds generally possess extended,
inhomogeneous envelopes rather than perfectly discontinuous edges, and the
location of a cloud boundary depends on the tracer and density threshold used
to define it \citep[e.g.,][]{2024ApJ...966..127M, 2013ApJ...763...51F}. Moreover, the dominant uncertainty in our treatment is not the
precise radial sharpness of the transition, but the assumption of spherical
symmetry itself. Real cloud breakout is intrinsically anisotropic and may
involve the separation of the ionization front from the dense swept-up shell,
photon leakage through porous structures, and directed champagne flows
\citep{1979A&A....71...59T,2019MNRAS.487.2200Z}. These
effects cannot be captured within a one-dimensional thin-shell model. We
therefore adopt the discontinuous profile as a minimal prescription that
preserves the uniform-cloud scalings used above while capturing the reduced
mass loading that follows expansion beyond the nominal cloud radius.

\subsection{Superbubble Formation}

In addition to breakout through the edge of the cloud, the H~II region may be
disrupted internally by the formation of a feedback-driven superbubble
\citep{1987ApJ...317..190M,1999MNRAS.309..332T,2013A&A...550A..49K}.
As the first core-collapse supernovae occur, their ejecta drive shocks into the
surrounding gas, sweeping the ambient medium into a dense shell and leaving
behind a hot, low-density cavity
\citep[e.g.,][]{trueloveEvolutionNonradiativeSupernova1999}. In a dense stellar
association, the remnants of multiple supernovae and stellar winds may overlap,
producing a collective superbubble that can become comparable to, or larger
than, the original H~II region.

Self-similar solutions for the evolution of wind- and supernova-driven bubbles were presented by \citet{1975ApJ...200L.107C,1977ApJ...218..377W}, and have since been widely applied to feedback-driven bubbles and superbubbles in star-forming environments \citep{1987ApJ...317..190M, 2004ApJ...605..751C, 1996ApJ...467..666O}.
In these models, a continuous mechanical luminosity, $L_{\rm mech},$ is deposited into an ambient medium of hydrogen number density $n_{\rm H}$, producing a hot ($\sim10^6$--$10^7\,{\rm K}$), shocked interior that drives a forward shock into the surrounding gas. 
The shocked ambient material cools efficiently and collapses into a thin, dense shell, whereas the much lower-density shocked wind and supernova ejecta remain hot, with radiative cooling times long compared with the bubble expansion time. The injected mechanical energy is therefore retained primarily as thermal energy in the interior, whose pressure drives the expansion of the radiative shell. We accordingly adopt the classical energy-driven thin-shell model throughout the bubble evolution (see, however, \citealt{2021ApJ...914...90L, 2021ApJ...914...89L}, which argue that the energy-driven model may over-estimate feedback efficiency). 

For analytical purposes, it is convenient to consider the evolution of the superbubble independent of the H~II region.
In this case, the problem is self-similar in nature and solutions have been found for the radius \citep{1977ApJ...218..377W},
\begin{equation}
    R_{\rm SB}(t)
    =
    0.76
    \left(\frac{L_{\rm mech} \Delta t^3}{m_p \mu_H n_{\rm H,0}}\right)^{1/5}.
    \label{eq:superbubble_radius}
\end{equation}
This solution is only approximate in the context of our model as the bubble will expand into the H~II region, which itself has a time-dependent density.
Additionally, the superbubble, like the H~II region, can eventually breakout of the cloud.

To provide a more complete treatment within our semi-analytic framework, we
evolve a pressure-driven superbubble shock propagating through the interior of
the H~II region. The superbubble is represented by a hot, approximately uniform
interior bounded by a thin swept-up shell. Its evolution is governed by a mechanical shock model \citep[e.g.,][]{beloborodovMechanicalModelRelativistic2006},
\begin{equation}
    \begin{aligned}
        \frac{dU_{\rm SB,int}}{dt}
        &=
        L_{\rm mech}
        -
        3\left(\gamma_{\rm ad}-1\right)
        \frac{U_{\rm SB,int}}{R_{\rm SB}}
        \frac{dR_{\rm SB}}{dt},
        \\
        \frac{dp_{\rm SB}}{dt}
        =
        4\pi R_{\rm SB}^{2}
        &\left[
            \left(\gamma_{\rm ad}-1\right)
            \frac{U_{\rm SB,int}}{V_{\rm SB}}
            -
            \frac{\mu_{\rm H}}{\mu_{\rm II}}
            n_{\rm H}(t)k_{\rm B}T_{\rm II}
        \right],
        \\
        \frac{dM_{\rm SB}}{dt}
        &=
        4\pi R_{\rm SB}^{2}
        \mu_{\rm H}m_p n_{\rm H}(t)
        \frac{dR_{\rm SB}}{dt},
        \\
        \frac{dR_{\rm SB}}{dt}
        &=
        \frac{p_{\rm SB}}{M_{\rm SB}},
    \end{aligned}
    \label{eq:superbubble_numerical_evolution}
\end{equation}
where $V_{\rm SB} = (4\pi/3) R_{\rm SB}^3$,  $U_{\rm SB,int}$ is the thermal energy of the hot interior, $\gamma_{\rm ad} = 5/3$ is the adiabatic constant, and $p_{\rm SB}=M_{\rm SB}\dot{R}_{\rm SB}$ is the radial momentum of the swept-up shell.
The density $n_{\rm H}(t)$ and pressure of the ambient ionized gas are obtained from the contemporaneous H~II-region solution, using Equation~\eqref{eq:dusty_shell_evolution} together with the implicit dusty ionization-balance condition in Equation~\eqref{eq:dust_stromgren_solution}.
We neglect the bulk velocity of the ionized gas when evaluating the superbubble mass-loading rate. 
This approximation is appropriate while the superbubble shock velocity substantially exceeds the characteristic expansion velocity of the H~II region.

Before the two shells meet, the H~II region and superbubble are evolved semi-independently: the H~II shell is driven by the pressure of the photoionized gas, while the superbubble shell is driven by the pressure of its hot interior and sweeps up the time-dependent ionized density.
We define the transition to the superbubble-dominated phase by $R_{\rm SB} = R_{\rm II}$.
At this time, the superbubble shell is assumed to overtake and merge with the photoionization-driven shell.
The masses and radial momenta of the two shells are combined, after which the common shell is evolved under the pressure of the hot superbubble interior.
The $10^4\,{\rm K}$ H~II-region pressure is then neglected as an independent source of dynamical support, and the interior is treated as a hot, dilute superbubble cavity.

The interaction between the superbubble and the ionizing radiation field is
necessarily more uncertain. 
The swept-up superbubble shell may cool rapidly and can, in principle, absorb ionizing photons before they reach the outer H~II shell.
A fully spherical treatment would therefore tend to reduce the ionizing luminosity available to maintain the H~II region once a dense inner shell forms.
Real superbubbles, however, are likely to be inhomogeneous and porous, with ionizing radiation escaping through low-density channels \citep{1989ApJ...337..141M, 1988ApJ...324..776M, 2017MNRAS.465.1720Y}.
Moreover, the shell requires a finite time to accumulate a sufficiently large absorbing column.
As a simplifying prescription, we therefore neglect attenuation by the superbubble shell while $R_{\rm SB}<R_{\rm II}$ and allow the full stellar ionizing luminosity to reach the H~II-region ionization front.
Once the shells merge, we terminate the volume-filling H~II-region treatment and instead calculate the ionized layer produced on the inner surface of the merged
superbubble shell.

The precise value of $L_{\rm mech}$ depends on the relative contributions of
stellar winds and supernovae, as well as on the efficiency with which the
injected energy is retained as thermal energy in the hot superbubble interior.
Each of these factors is uncertain and subject to debate in the literature \citep[e.g.,][]{2004ApJ...605..751C,1996ApJ...467..666O, 2018A&A...615A..40R, 2011ApJ...731...91L, 2017MNRAS.468.2757V, 2019MNRAS.490.1961E}.
As a fiducial estimate, we neglect the contribution from stellar winds and
consider only the energy input from core-collapse supernovae, which dominate
the mechanical feedback after the first few Myr of stellar evolution.

We adopt an effective core-collapse progenitor mass interval
$M_{\rm SN}=8$--$40\,M_\odot$ \citep{heger_how_2003,2015PASA...32...16S,2009ARA&A..47...63S}. 
For a \citet{2001MNRAS.322..231K} initial mass function normalized per unit stellar
mass formed, the number of core-collapse progenitors per unit stellar mass is therefore
\begin{equation}
    \zeta_{\rm SN}
    \equiv
    \frac{N_{\rm SN}}{M_\star}
    =
    \int_{8\,M_\odot}^{40\,M_\odot} \xi(M)\,dM
    \simeq
    10^{-2}\,M_\odot^{-1}.
\end{equation}
The most massive stars in this interval undergo core collapse after
$t_{\rm first}\simeq 3$--$4\,{\rm Myr}$, while the least massive progenitors
explode after $t_{\rm last}\simeq 30$--$50\,{\rm Myr}$. We therefore define
\begin{equation}
    \Delta t_{\rm SN}
    \equiv
    t_{\rm last}-t_{\rm first},
\end{equation}
and adopt $\Delta t_{\rm SN}=37\,{\rm Myr}$ as a representative duration for
the supernova-feedback epoch.
If each supernova releases a mechanical energy $E_{\rm SN}\simeq 10^{51}\,{\rm erg}$ \citep{2021Natur.589...29B}, the time-averaged supernova power is
\begin{equation}
    \begin{aligned}
    L_{\rm SN}
    &\simeq
    \frac{\zeta_{\rm SN} M_\star E_{\rm SN}}{\Delta t_{\rm SN}},\\
    L_{\rm SN}
    &\simeq
    8.6\times10^{39} \;{\rm erg\,s^{-1}}\; E_{\rm SN, 51}
    M_{\star,6}
    \end{aligned}
\end{equation}
Not all of this power is retained in the hot bubble. Radiative cooling,
turbulent mixing, leakage through low-density channels, and incomplete coupling
to the surrounding gas can reduce the effective luminosity that drives the
superbubble expansion. We therefore write
\begin{equation}
    L_{\rm mech}
    =
    \epsilon_{\rm mech} L_{\rm SN},
\end{equation}
where $\epsilon_{\rm mech}\leq 1$ is an effective mechanical-energy retention
efficiency. 
We adopt a fiducial value $\epsilon_{\rm mech} = 0.1$ \citep{2014MNRAS.443.3463S,2017MNRAS.465.1720Y,2018MNRAS.481.3325F}.

Using this fiducial closure for the mechanical luminosity, the superbubble radius is (assuming equation~\ref{eq:superbubble_radius} to be valid)
\begin{equation}
    \label{eq:superbubble_radius_scaling}
    R_{\rm SB} \simeq 61\;{\rm pc}\;\epsilon_{\rm mech, -1}^{1/5} E_{\rm SN, 51}^{1/5} M_{\star,6}^{3/10} \eta_{-1}^{-1/10} \Sigma_{\rm cl,2}^{-3/10} \Delta t_{6}^{3/5}.
\end{equation}
As a fraction of the cloud radius, this is
\begin{equation}
    \frac{R_{\rm SB}}{R_{\rm cl}}
    \simeq
    0.34\;\epsilon_{\rm mech, -1}^{1/5} E_{\rm SN, 51}^{1/5} M_{\star,6}^{-1/5} \eta_{-1}^{2/5} \Sigma_{\rm cl,2}^{1/5} \Delta t_{6}^{3/5},
\end{equation}
and the crossing time will be 
\begin{equation}
    t_{\rm SB, cross}
    \simeq
    6\;{\rm Myr}\; \epsilon_{\rm mech, -1}^{-1/3} E_{\rm SN, 51}^{-1/3} M_{\star,6}^{1/3} \eta_{-1}^{-2/3} \Sigma_{\rm cl,2}^{-1/3}.
\end{equation}
Thus, for the fiducial parameters, the superbubble can reach the edge of the
parent cloud on a timescale comparable to or shorter than the H~II-region
breakout time. 

At early times, when $R_{\rm SB}\ll R_{\rm II}$, the superbubble occupies only
the central part of the H~II region and the FRB sightline is still dominated by
the volume-filling ionized gas described above.
Once $R_{\rm SB}\sim R_{\rm II}$, however, the smooth H~II-region approximation is no longer self-consistent.
The superbubble interior is hot and dilute, while the swept-up gas is concentrated into a dense shell whose contribution to the DM depends on its geometry, compression, and ionization state. 

\begin{figure*}[ht]
    \resizebox{\hsize}{!}{
    \includegraphics[width=1\linewidth]{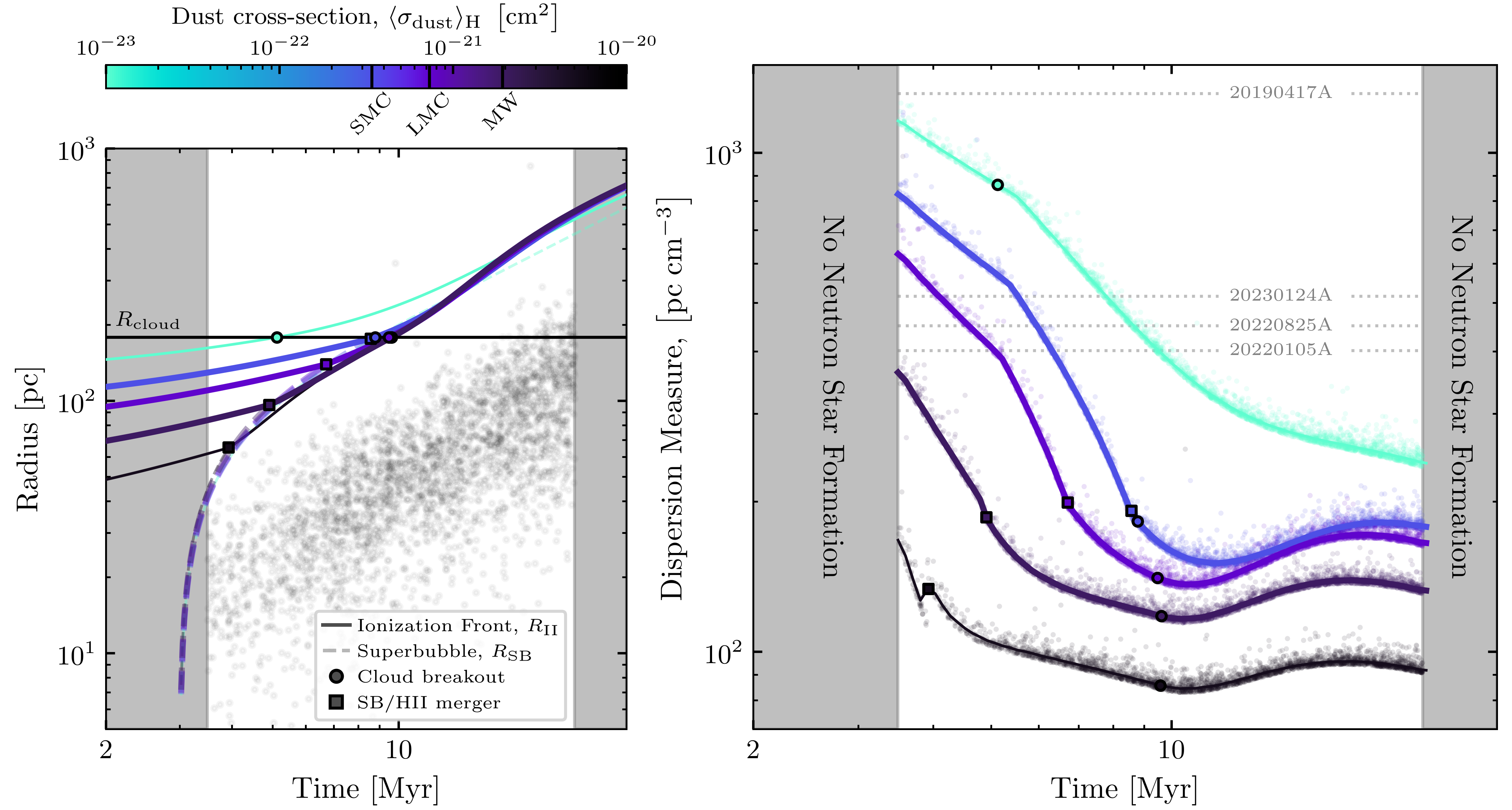}
   }
\caption{
Dependence of the semi-analytic, constant ionizing flux FRB DM model on the effective
Lyman-continuum dust absorption cross section, $\langle \sigma_{\rm dust}\rangle_{\rm H}$.
The left panel shows the evolution of the H\,II-region radius, $R_{\rm II}(t)$
(solid), and superbubble radius, $R_{\rm SB}(t)$ (dashed), for a range of dust
opacities, with curve color corresponding to $\langle \sigma_{\rm dust}\rangle_{\rm H}$ as indicated by
the color bar. The horizontal black line marks the parent-cloud radius,
$R_{\rm cloud}$. Circles mark the cloud-breakout time (see section~\ref{sect:cloud_breakout}) and squares mark the
SB/H\,II merger time for each model, after which the dashed $R_{\rm SB}$
curve is replaced by the merged shell radius. All simulations used a fiducial $M_\star = 10^6\;{\rm M_\odot}$.
The right panel shows the dispersion measures of the corresponding Monte
Carlo FRB sightlines as a function of activation time, colored by
$\langle \sigma_{\rm dust}\rangle_{\rm H}$ as in the left panel. The source-frame dispersion measures of a subset of high-DM FRBs (20190417A, 20230124A, 20220825A, and 20220105A) are indicated for comparison using the data of \citet{2026ApJ...997L...6Y}. See section~\ref{sect:observational_data} for a detailed description of the sample. 
}
\label{fig:dust_progression}
\end{figure*}

To construct a heuristic model for the superbubble contribution to the dispersion measure, we consider the FRB sightline through the swept-up shell driven by the expanding superbubble. 
Conservation of mass requires that
\begin{equation}
    \Sigma_{\rm H,sh}
    \equiv
    n_{\rm H,sh}\Delta R_{\rm sh}
    =
    \frac{n_{\rm H,0}R_{\rm SB}}{3},
    \label{eq:sb_total_column}
\end{equation}
independent of the unknown physical shell thickness $\Delta R_{\rm sh}$.
In the classical energy-driven solution, the pressure of the hot
superbubble interior can be written as
\begin{equation}
    P_{\rm SB,int}
    =
    (\gamma_{\rm ad} -1) \frac{U_{\rm SB,int}}{V_{\rm SB}}.
    \label{eq:sb_interior_pressure}
\end{equation}
Ionizing photons from the central stellar population illuminate the inner
surface of the swept-up shell, producing an ionized skin with temperature
$T_{\rm II}$. As a heuristic closure, we assume that this ionized layer is
in pressure equilibrium with the hot interior. Its hydrogen-nucleus density
is therefore
\begin{equation}
    n_{\rm H,ion}
    =
    \frac{\mu_{\rm II}}{\mu_{\rm H}}
    \frac{P_{\rm SB,int}}{k_{\rm B}T_{\rm II}}
    =
    \frac{\mu_{\rm II}}{\mu_{\rm H}} (\gamma_{\rm ad} - 1) \frac{U_{\rm SB,int}}{k_B T_{\rm II} V_{\rm SB}}
\end{equation}
Here we assume that hydrogen is fully ionized within the skin, such that
$n_e\simeq n_p\simeq n_{\rm H,ion}$.

Approximating the ionized layer as locally plane parallel, the dusty photon
budget gives the formal ionized depth
\begin{equation}
    \Delta R_{\rm ion}
    =
    \frac{1}{n_{\rm H,ion}\langle \sigma_{\rm dust}\rangle_{\rm H}}
    \ln\left[
        1+
        \frac{\langle \sigma_{\rm dust}\rangle_{\rm H}\mathbb{N}_{\rm Lyc}}
        {4\pi\alpha_B(T_{\rm II})
         n_{\rm H,ion}R_{\rm SB}^{2}}
    \right].
    \label{eq:sb_ionized_depth}
\end{equation}
In the dust-free limit, this reduces to
\begin{equation}
    \Delta R_{\rm ion}
    =
    \frac{\mathbb{N}_{\rm Lyc}}
    {4\pi\alpha_B(T_{\rm II})
     n_{\rm H,ion}^{2}R_{\rm SB}^{2}}.
\end{equation}

Because Equation~\eqref{eq:sb_ionized_depth} does not explicitly account
for the finite amount of swept-up material, the radial ionized column is
limited to the total shell column given by
Equation~\eqref{eq:sb_total_column}. The superbubble contribution along a
radial sightline crossing one side of the shell is therefore
\begin{equation}
    {\rm DM}_{\rm SB}^{\perp}
    =
    \min\left[
        n_{\rm H,ion}\Delta R_{\rm ion},
        \frac{n_{\rm H,0}R_{\rm SB}}{3}
    \right].
    \label{eq:sb_radial_dm}
\end{equation}
For sightlines that intersect the shell at an oblique angle, the path length through the ionized layer is longer by a factor of $1/\cos\theta$, where $\theta$ is the angle between the sightline and the local shell normal.

To obtain a continuous dispersion-measure prescription across the transition
from the H~II-region-dominated phase to the superbubble-dominated phase, we
assume that the hot superbubble interior is sufficiently dilute to contribute
negligibly to the electron column. Before the two shells merge, the total
dispersion measure is therefore written as
\begin{equation}
    {\rm DM}_{\rm tot}
    =
    {\rm DM}_{\rm SB}
    +
    {\rm DM}_{\rm HII,res},
    \qquad
    R_{\rm SB}<R_{\rm II},
    \label{eq:dm_sb_pre_crossing}
\end{equation}
where ${\rm DM}_{\rm SB}$ is the contribution from the ionized inner layer of
the superbubble shell and ${\rm DM}_{\rm HII,res}$ is the contribution from
the residual volume-filling ionized gas lying between the superbubble shell
and the outer H~II-region boundary. This construction avoids double counting
the gas that has already been swept into the superbubble shell.

For a source near the center of the association and a radial sightline, the
residual H~II-region contribution is approximately
\begin{equation}
    {\rm DM}_{\rm HII,res}
    =
    n_{\rm H}(t)
    \left[
        R_{\rm II}(t)-R_{\rm SB}(t)
    \right].
    \label{eq:dm_hii_residual}
\end{equation}
For off-center sources and arbitrary viewing directions, we instead evaluate
the electron column through the portion of the H~II region exterior to
$R_{\rm SB}$ using the corresponding difference between the outer and inner
spherical chord lengths.

As $R_{\rm SB}\rightarrow R_{\rm II}$, the residual volume-filling path
length vanishes continuously, such that
\begin{equation}
    {\rm DM}_{\rm HII,res}\rightarrow0.
\end{equation}
Once the superbubble shell overtakes and merges with the
photoionization-driven shell, we terminate the volume-filling H~II-region
contribution and adopt
\begin{equation}
    {\rm DM}_{\rm tot}
    =
    {\rm DM}_{\rm SB},
    \qquad
    R_{\rm SB}\geq R_{\rm II}.
    \label{eq:dm_sb_post_crossing}
\end{equation}
The mass and momentum of the photoionization-driven shell are incorporated
into the merged superbubble shell at this transition, and its ionized column
is subsequently calculated from the ionized-skin prescription described
above. This closure neglects the detailed hydrodynamic interaction between the
two shells, but provides a continuous interpolation between the
volume-filling H~II-region phase and the later shell-dominated phase.

The closure presented here for the superbubble dominated phases of the evolution
should be understood as a heuristic model. 
Interactions between the wind-driven expansion of the superbubble, the expansion of the H~II region, and
the eventual breakout of either or both from the parent cloud are complex and highly geometry dependent.
We therefore defer a more detailed treatment of these effects to future work, and we focus here on
the early-time evolution of the H~II region, which is well described by the analytic model presented
in Section~\ref{sect:modeling}.
The heuristic treatments presented in this section are, however, sufficient to demonstrate the critical insight:
the DM contribution from the H~II region is largest at early times, when the ionized gas is dense and the path length is long, 
and it declines as the H~II region expands and the density drops.

\subsection{Monte Carlo Realization}
\label{sec:monte_carlo_analytic}

We realize the semi-analytic model using a Monte Carlo sampling procedure.
For each stellar association, we draw FRB progenitor masses from the adopted
IMF over the neutron-star-producing mass range and assign each progenitor a
core-collapse time using the mass--lifetime relation in
Equation~\eqref{eq:stellar_lifetime}. Initial progenitor positions are
drawn from the Plummer profile describing the stellar association, with
isotropic angular coordinates.

Each progenitor is displaced from its birth site by a main-sequence drift
$\Delta\mathbf{r}_{\rm MS}$, drawn as a three-dimensional Gaussian with
independent, identically distributed Cartesian components,
\begin{equation}
    \Delta {\bf r}_{{\rm MS}} \sim
    \mathcal{N}\!\left(\boldsymbol{0},\,\sigma_{\rm MS}^2\mathbb{I}_3\right),
    \qquad
    \sigma_{\rm MS} = \frac{v_{\rm drift}\,\tau_{\rm MS}(M)}{\sqrt{3}}.
\end{equation}
At core collapse, we additionally draw an isotropic natal-kick velocity
$\mathbf{v}_{\rm kick}$, with independent, identically distributed
Cartesian components,
\begin{equation}
    \mathbf{v}_{\rm kick} \sim
    \mathcal{N}\!\left(\mathbf{0},\,\sigma_{\rm kick}^2\mathbb{I}_3\right),
\end{equation}
and propagate the newly formed neutron star for a post-collapse delay
$\Delta t_{\rm FRB}$ drawn from the activity kernel in
Equation~\eqref{eq:psi_prompt}, giving a kick displacement
\begin{equation}
    \Delta\mathbf{r}_{\rm kick} = \mathbf{v}_{\rm kick}\,\Delta t_{\rm FRB}.
\end{equation}
The FRB activation time is therefore
\begin{equation}
    t_{\rm FRB} = t_{\rm cc} + \Delta t_{\rm FRB},
\end{equation}
and the source position at activation is
\begin{equation}
    \mathbf{r}(t_{\rm FRB}) = \mathbf{r}_{\rm birth} + \Delta\mathbf{r}_{\rm MS} + \Delta\mathbf{r}_{\rm kick}.
\end{equation}

Due to the spherical symmetry of the system, we assume (without loss of generality) that the line-of-sight (LOS) to each synthetic source is along the $+z$ axis, with the observer at $z\to+\infty$. For a source at $(x,y,z_s)$ with impact parameter $b=\sqrt{x^2+y^2}$
relative to the center of the association, the forward chord length through
a sphere of radius $R$ centered on the origin is
\begin{equation}
    \ell(R) =
    \begin{cases}
        \max \left[0, z_{+} - \max\left(z_s,\,z_{-}\right)\right], & b<R,\\
        0, & b\geq R,
    \end{cases}
    \label{eq:chord_length}
\end{equation}
with $z_{\pm}(R)=\pm\sqrt{R^2-b^2}$. Equation~\eqref{eq:chord_length}
correctly reduces to the full forward chord for sources outside the sphere on
the near side, the truncated forward path length for embedded sources, and
zero once the source lies beyond the sphere's far edge.

Because $\ell(R)$ is additive over nested spheres, the residual,
volume-filling H~II region contribution, the annulus between the outer
ionized radius $R_{\rm II}$ and the (possibly zero, pre-superbubble) inner
cavity radius $R_{\rm SB}$, is obtained directly as
\begin{equation}
    {\rm DM}_{\rm res}
    =
    n_{\rm H}\bigl[\ell(R_{\rm II}) - \ell(R_{\rm SB})\bigr],
    \label{eq:dm_res_general}
\end{equation}
which generalizes Equation~\eqref{eq:dm_hii_residual} to arbitrary viewing
geometry and reduces to it for a radial sightline from the center. The
swept-up superbubble shell contributes an ionized skin of density
$n_{\rm H,ion}$ and physical depth $\Delta R_{\rm ion}$
(Equation~\eqref{eq:sb_ionized_depth}) on its inner face, evaluated the same
way as a thin spherical annulus,
\begin{equation}
    {\rm DM}_{\rm SB}^{\rm raw}
    =
    n_{\rm H,ion}
    \bigl[
        \ell(R_{\rm SB}) - \ell(R_{\rm SB}-\Delta R_{\rm ion})
    \bigr].
    \label{eq:dm_skin_raw}
\end{equation}

Because Equation~\eqref{eq:dm_skin_raw} does not by itself guarantee that the
implied column does not exceed the shell's total swept-up material, we cap it
using the exact areal hydrogen column obtained from the shell's
tracked swept-up mass $M_{\rm SB}$,
\begin{equation}
    \Sigma_{\rm H,sh}
    =
    \frac{M_{\rm SB}}{4\pi R_{\rm SB}^2\,\mu_{\rm H}m_p},
    \label{eq:sigma_H_sh_exact}
\end{equation}
projected along the sightline by the appropriate obliquity factor. For a ray
crossing a geometrically thin sphere of radius $R$ at impact parameter $b$,
each crossing contributes a secant factor $R/z_{\rm far}(R)$, and the number
of forward crossings is 0 (source beyond the shell, $z_s\geq z_{\rm far}$), 1
(source already inside the shell, $z_{\rm near}\leq z_s<z_{\rm far}$), or 2
(source has not yet reached the shell, $z_s<z_{\rm near}$). Denoting this
crossing-weighted secant factor $\kappa(R_{\rm SB})$, the shell contribution
is
\begin{equation}
    {\rm DM}_{\rm SB}
    =
    \min\!\left(
        {\rm DM}_{\rm SB}^{\rm raw},\
        \Sigma_{\rm H,sh}\,\kappa(R_{\rm SB})
    \right),
    \label{eq:dm_skin_capped}
\end{equation}
which generalizes the simpler oblique-angle prescription of
Equation~\eqref{eq:sb_radial_dm} to sightlines that may cross the shell zero,
one, or two times.

The total dispersion measure along a given sightline is then
${\rm DM}={\rm DM}_{\rm res}+{\rm DM}_{\rm SB}$, evaluated using the source's
phase at $t_{\rm FRB}$: for H~II-only sources $R_{\rm SB}=0$ and
${\rm DM}_{\rm SB}$ vanishes identically; for merged sources $R_{\rm
II}=R_{\rm SB}$ by construction and ${\rm DM}_{\rm res}$ vanishes identically,
leaving ${\rm DM}={\rm DM}_{\rm SB}$ alone, consistent with
Equation~\eqref{eq:dm_sb_post_crossing}. 

Repeating this procedure over the full Monte Carlo population yields a
phase-dependent realization of the dispersion-measure distribution for the
fiducial model. Figure~\ref{fig:dust_progression} illustrates the results of
this procedure for a range of dust cross sections,
$\langle\sigma_{\rm dust}\rangle_{\rm H}$, showing how the H\,II-region and
superbubble evolution, and the resulting DM distribution, depend on dust
opacity.

\section{Idealized Numerical Simulations}
\label{sect:numerical}

The semi-analytic model developed in Section~\ref{sect:modeling} assumes a constant ionizing luminosity. While this approximation is adequate at early times, it breaks down once the most massive stars begin to leave the main sequence and undergo core collapse, causing the Lyman-continuum output of the population to decline rapidly. Because this is also the epoch during which the first FRB-capable neutron stars are expected to form, the resulting time dependence of the ionizing radiation field is central to the predicted dispersion measure.

Extending the semi-analytic model to this regime is not straightforward. As the ionizing luminosity fades, the ionization front can recede or separate from the forward shock, invalidating the thin-shell assumption that $R_{\rm II}=R_{\rm FS}$. We therefore resort to idealized one-dimensional radiation-hydrodynamic (RHD) simulations to more completely capture the relevant physics.

\subsection{Numerical Method}

During the hydrodynamic step, the solver evolves the spherically symmetric gas
density and radial velocity according to the inviscid Euler equations in spherical coordinates,
\begin{align}
    \frac{\partial \rho}{\partial t} + \frac{1}{r^2} \frac{\partial}{\partial r} \left(r^2 \rho v\right) &= 0,\\
    \frac{\partial(\rho v)}{\partial t} + \frac{1}{r^2}\frac{\partial}{\partial r}\left[r^2\left(\rho v^2 + P\right)\right] &= \frac{2P}{r}.
\end{align}
The equations are discretized using a first-order finite-volume Godunov scheme, with HLL approximate Riemann fluxes evaluated at cell interfaces.

Between hydrodynamic updates, the radiation step recomputes the
ionization structure using the on-the-spot approximation
\citep[e.g.,][]{draine2011physics,2005pcim.book.....T}. Ionizing photons are
propagated outward from the central stellar population and attenuated by both
hydrogen recombinations and dust absorption according to
Equation~\eqref{eq:dusty_photon_budget}.

From the resulting ionization profile, $x(r,t)$, we compute the gas
pressure using a two-temperature equation of state
\begin{equation}
    P =
    \frac{\rho k_{\rm B}}{\mu_{\rm I}m_{\rm H}}
    \left[
        \frac{\mu_{\rm I}}{\mu_{\rm II}}xT_{\rm II}
        +(1-x)T_{\rm I}
    \right],
\end{equation}
where $T_{\rm I}$ and $T_{\rm II}$ are the neutral- and ionized-gas
temperatures, respectively, and $\mu_{\rm I}$ and $\mu_{\rm II}$ are the
corresponding mean molecular weights. This prescription captures the pressure
contrast that drives the expansion of the H~II region while avoiding the need
to solve a full thermal-energy equation. In all simulations presented here, we take $T_{\rm I} = 10\;{\rm K}$, but note that our results are robust to any choice of $T_{\rm I}$ so long as it is not low enough to cause numerical overflow, or high enough to spuriously introduce an antagonistic thermal pressure source against the evolution of the H~II region. Further details of the numerical
scheme are given in Appendix~\ref{app:numerical_method}.

\subsection{Initial Conditions and Boundary Conditions}

The simulations begin from a static, initially neutral cloud with uniform
hydrogen number density $n_{\rm H,0}$ within radius $R_{\rm cl}$, determined from
Equations~\eqref{eq:n_H_scale} and \eqref{eq:R_cl_scaling}. Beyond the cloud radius, the density is set to a non-zero floor representing the ambient ISM (see equation~\ref{eq:cloud_ism_density_profile}). The initial radial
velocity is set to zero throughout the domain. 
The radial domain is discretized logarithmically from a finite inner radius,
$r_{\rm in}>0$, to an outer radius $r_{\rm out} > 0$. 
An additional central cell is added from $r = 0$ to the inner boundary.
We impose a reflective boundary at $r=0$, suitable for spherical symmetry, and impose an outflow boundary at $r_{\rm out}$.
The outer boundary is placed well beyond $R_{\rm cl}$ so that the H~II region can expand through the cloud edge without interacting with the edge of the computational domain.

The stellar population is treated as a central source of ionizing radiation.
The ionizing photon rate entering the innermost radial cell is set by the
instantaneous stellar-population luminosity,
$\mathbb{N}_{\rm Lyc}(t)$. The radiation field is then propagated outward
through the grid, and the ionization state is updated from the inner boundary
until the available photon budget is exhausted.

Figure~\ref{fig:radhd_comparison} compares the resulting ionization-front
evolution with the modified Spitzer solution and the thin-shell model developed
in Section~\ref{sect:modeling}. For constant ionizing luminosity, the RHD
calculation closely reproduces the expected pressure-driven expansion of the
H~II region. The thin-shell solution provides the closest agreement, while the
modified Spitzer approximation captures the overall evolution with modest
deviations at late times and high dust opacity. This comparison verifies that
the numerical scheme recovers the standard H~II-region limit before it is
applied to a time-dependent stellar radiation field.

\begin{figure}
    \centering
    \includegraphics[width=\linewidth]{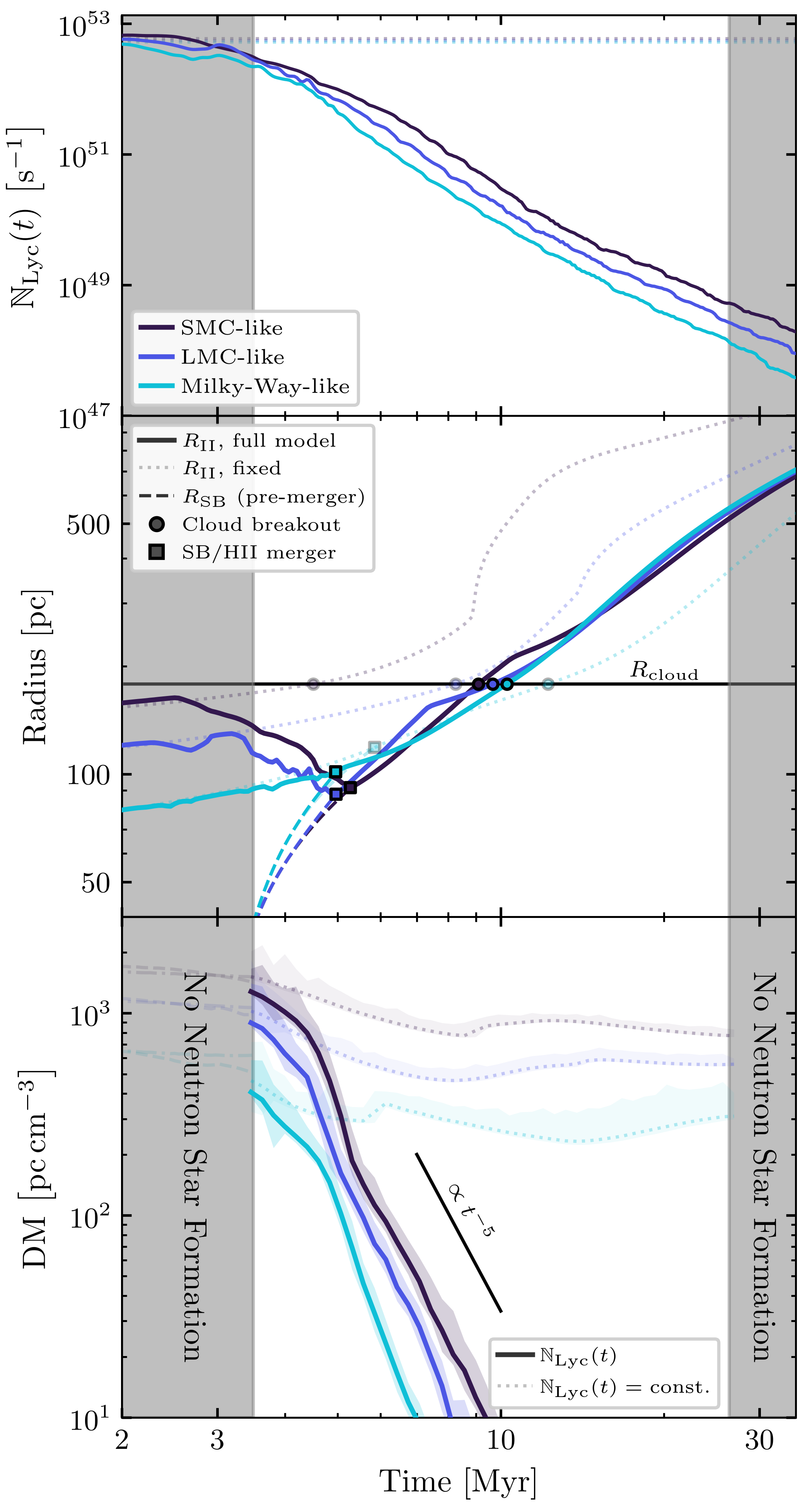}
    \caption{
    Comparison of the ionizing luminosity, ionized-region radius, and median DM for a fiducial $M_\star = 10^6\;{\rm M_\odot}$ stellar association at SMC-, LMC-, and Milky-Way-like metallicities.
    From top to bottom, the panels show the ionizing photon production rate, $\mathbb{N}_{\rm Lyc}(t)$, the radii of the ionization front and superbubble, and the DM distribution of synthetic FRBs.
    Solid lines correspond to time-dependent $\mathbb{N}_{\rm Lyc}$, while dotted lines show cases held fixed at the same initial value.
    In the middle panel, the horizontal line marks the cloud radius, circles indicate cloud breakout, and squares indicate merger of the superbubble with the H~II-region boundary.
    In the lower panel, the dark black line indicates the best-fit power-law slope to the ${\rm DM}$ decay ($t^{-5}$).
    Gray regions denote times outside the neutron-star formation interval.
    During this interval, DMs are obtained from the Monte Carlo population model; at earlier times, the plotted value is the effective central DM.
    }
    \label{fig:time_dependence}
\end{figure}

\subsection{Stellar Population Synthesis and Dust Absorption}

Ionizing photon production rates and stellar spectra are computed using \texttt{pySTARBURST99} \citep{2025ApJS..280....5H}, assuming a \citet{2001MNRAS.322..231K} initial mass function. We generate spectra for a fiducial stellar population mass $M_\star=10^6\,M_\odot$ and rescale the resulting luminosities linearly to the stellar population mass used in each simulation. The \texttt{pySTARBURST99} grid includes six metallicity templates: zero metallicity, IZw18-like ($Z=0.0004$), SMC-like ($Z=0.002$), LMC-like ($Z=0.006$), Milky-Way-like ($Z=0.014$), and a super-solar Galactic-center-like model ($Z=0.02$). In this work, we focus on four representative environments: IZw18-like, SMC-like, LMC-like, and Milky-Way-like stellar populations. These span the range from extremely metal-poor blue compact dwarf galaxies to approximately solar-metallicity star-forming regions.
For metallicities that do not coincide with one of these four templates, we interpolate the ionizing flux between the two nearest bracketing metallicity models.

Dust absorption of Lyman-continuum photons is included through effective absorption cross sections per hydrogen nucleus derived from the carbonaceous--silicate grain models of \citet{2001ApJ...548..296W}. 
From the extinction cross section per hydrogen ($C_{\rm ext}/H$) and albedo ($\omega$) provided by the dust models, we compute the effective dust absorption cross section as
\begin{equation}
    \sigma_{\rm dust\;H,\lambda} = (1-\omega_\lambda)\left(\frac{C_{\rm ext, \lambda}}{\rm H}\right).
\end{equation}
The factor $(1-\omega_\lambda)$ converts the total extinction cross section to an effective absorption cross section. In our photon-number treatment, scattering is taken to redirect rather than destroy ionizing photons and is therefore not included as a direct sink.

The effective cross section used in the numerical simulations was then computed as the photon-number-weighted mean using the ionizing SED modeled by \texttt{pySTARBURST99},
\begin{equation}
    \langle \sigma_{\rm dust}\rangle_{\rm H}(t)
    =
    \frac{
    \displaystyle
    \int_{\lambda_{\min}}^{\lambda_{\rm Lyc}}
    \sigma_{\rm dust\;H,\lambda}
    \left[
    \frac{L_\lambda(t)\lambda}{hc}
    \right]\,d\lambda
    }{
    \displaystyle
    \int_{\lambda_{\min}}^{\lambda_{\rm Lyc}}
    \left[
    \frac{L_\lambda(t)\lambda}{hc}
    \right]\,d\lambda
    }.
    \label{eq:effective_dust_cross_section}
\end{equation}
While equation~\eqref{eq:effective_dust_cross_section} produces a time-dependent cross section $\langle \sigma_{\rm dust}\rangle_{\rm H}(t)$, the resulting cross section varies by at most 1.5\% over the $\sim 10\;{\rm Myr}$ timescale of interest.
We therefore adopt the mean value at all times.

We use the Milky-Way, LMC, and SMC dust models as empirical proxies for environments with comparable metallicities and extinction properties.
For metallicities that do not coincide with these tabulated values, we interpolate the effective dust absorption cross section log-linearly in $\log_{10} Z$ between the two nearest bracketing models.
For metallicities outside the tabulated range, we extrapolate using the slope defined by the nearest pair of templates.
This prescription provides a practical estimate of the Lyman-continuum dust opacity, but it does not capture variations in grain composition, size distribution, or dust-to-gas ratio that may occur among galaxies with similar metallicities.
Figure~\ref{fig:time_dependence} compares the full time-dependent calculation with the corresponding fixed-luminosity models. The decline of $\mathbb{N}_{\rm Lyc}(t)$ during the neutron-star formation epoch substantially reduces both the extent of the ionized region and the dispersion measure available to embedded FRB sources at late times.

\subsection{Superbubble Evolution}

Although the two-temperature equation of state and the assumption of ionization equilibrium make the RHD calculation quite simple, they do not provide the thermal-energy evolution needed to model the hot superbubble interior self-consistently.
We therefore defer a fully coupled treatment of photoionization and mechanical feedback to future work and instead incorporate the superbubble in post-processing using a similar method to the one introduced in the semi-analytic model.

The superbubble is evolved using Equation~\eqref{eq:superbubble_numerical_evolution}, with the ambient ionized-gas density supplied by the RHD simulation. Specifically, at each time we evaluate the instantaneous density profile, $n_{\rm H}(r,t)$, at the superbubble shock radius and use this value to determine the mass loading and external pressure entering the superbubble evolution.

\subsection{FRB Dispersion Measures}

Following Section~\ref{sect:modeling}, we generate a synthetic FRB population and place each source within the simulation domain. For each source, the ionization-front radius and radial electron-density profile are evaluated at its activation time, $t_{\rm FRB}$, by linearly interpolating between the two bracketing simulation snapshots.

The residual (H~II) dispersion measure is computed directly from the
simulated radial electron-density profile. For
$b < R_{\rm II}(t_{\rm FRB})$, define
$z_\pm = \pm [R_{\rm II}^2(t_{\rm FRB})-b^2]^{1/2}$.
For sources with $z_s < z_+$,
\begin{equation}
    {\rm DM}_{\rm HII,res}
    =
    \int_{\max(z_s,z_-)}^{z_+}
    n_e\!\left(\sqrt{b^2+z^2},t_{\rm FRB}\right)\,dz,
    \label{eq:simulation_dm_los}
\end{equation}
where we modify $n_e$ such that $n_e(r,t)=0$ for $r<R_{\rm SB}(t)$ to account for the loss of volume due to superbubble expansion.
If $z_s \ge z_+$, ${\rm DM}_{\rm HII,res}=0$.
We evaluate this integral using Gauss--Legendre quadrature. At each quadrature
point, the electron density is obtained by linear interpolation in radius,
while the full radial profile is interpolated in time between the two
simulation snapshots surrounding $t_{\rm FRB}$.
The superbubble's contribution to the DM is computed using the same procedure as described in section~\ref{sect:modeling}.

\begin{figure*}[ht]
    \resizebox{\hsize}{!}{
    \centering
    \includegraphics[width=1\linewidth]{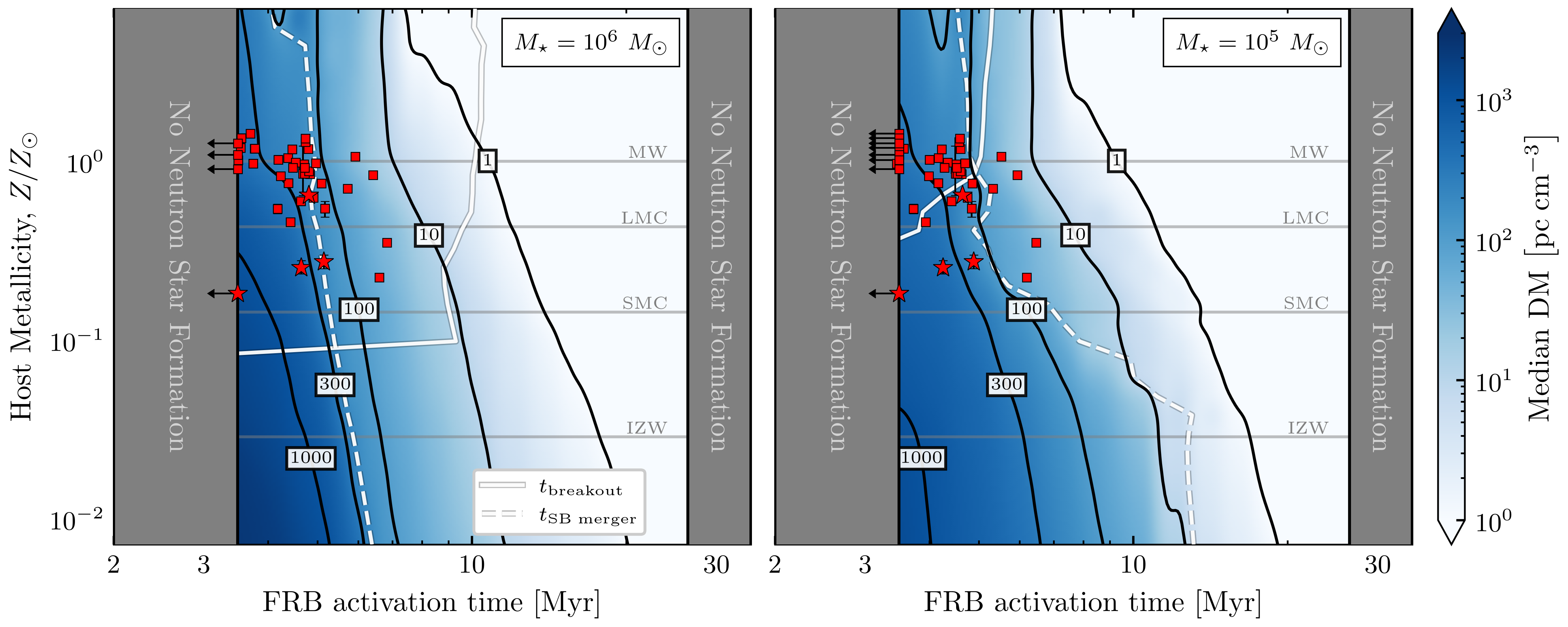}
    }
    \caption{
    Median dispersion measure of synthetic FRBs as a function of host metallicity and FRB activation time for stellar associations with
    $M_\star = 10^6\;{\rm M_\odot}$ (left) and $10^5\;{\rm M_\odot}$ (right).
    Black contours mark constant-DM levels, while the solid and dashed white curves indicate the superbubble-merger and cloud-breakout timescales, respectively.
    Horizontal gray lines mark the metallicities of the adopted stellar-population templates.
    Red points show the volume-limited FRB sample of \citet{2026ApJ...997L...6Y}.
    For each source, the activation time is inferred by matching its observed host-frame DM to the simulated DM at the corresponding host metallicity.
    In cases where a source could not be matched to an activation time (i.e. the DM exceeded the maximum produced in fiducial simulations), they are indicated with left-facing arrows.
    Stars mark the four FRBs associated with persistent radio sources (PRSs) identified by \citet{moroianuMilliarcsecondLocalizationAssociates2025}.
    }
    \label{fig:RHD_fiducial}
\end{figure*}

\section{Observational Data}
\label{sect:observational_data}

To facilitate comparison between our model and observed FRBs, we compile DM measurements for key sources with known host metallicities \citep{2026ApJ...997L...6Y}; these sources are shown in figures~\ref{fig:dust_progression} and \ref{fig:RHD_fiducial}.
Uncertainties are estimated using values for $\sigma_\mathrm{cosmic}$ estimated from~\citet{konietzka2025}, neglecting the typical $\sim 20\%$ uncertainties from the subdominant Galactic DM contribution. 
DM excesses for FRB 20121102 are estimated in~\citet{tendulkar2017host}, who infer an observed-frame range of $50-250\;{\rm pc/cm^3}$ for the host contribution.
For FRB 20190417A,~\citet{moroianuMilliarcsecondLocalizationAssociates2025} reports a rest-frame excess of $1275 \;{\rm pc/cm^3}$ ($>1228 \;{\rm pc/cm^3}$ at 90\% confidence).
For FRB 20190520B and 20240114A, large DM excesses were initially reported~\citep{niuRepeatingFastRadio2022}. 
However, the discovery of an intervening galaxy cluster~\citep{2023ApJ...954L...7L} revised DM excesses to $210-570 \;{\rm pc/cm^3}$ for FRB 20190520B and $100-260 \;{\rm pc/cm^3}$ for 20240114A; these uncertainties, however, do not reflect intrinsic uncertainties in e.g. the cluster gas profiles.

FRB 20250613A, which has a very \emph{low} rest-frame DM excess of $52 \pm 40 \;{\rm pc/cm^3}$, provides compelling evidence for the existence of long-delay FRB activation.
The host has a similar stellar mass ($\log_{10} M_\ast = 8.31$) and low metallicity ($Z/Z_\odot \approx 0.22$) to the hosts in the sample, but unlike the others, the host-integrated star formation history suggests a quenching event $\sim 25$ Myr ago rather than ongoing star formation~\citet{dial2026}.
The natural explanation for this source is that the DM excess is low because of the long delay since the most recent starburst, during which DM contributions powered by massive star formation scales have decayed.
This source shows that low host stellar mass and gas-phase metallicities are alone insufficient to generate large DM excesses, demonstrating the causal role played by star formation in our model.

\section{Results} 
\label{sect:results} 

\subsection{A Model for Ultra-High Dispersion FRBs}

This work has demonstrated that young massive stellar associations can produce H~II-region electron columns
large enough to account for the source-frame dispersion measures of the
low-redshift, high-DM FRBs identified by
\citet{moroianuMilliarcsecondLocalizationAssociates2025}
\citep[see also, e.g.,][]{bhandariNonrepeatingFastRadio2023,
niuRepeatingFastRadio2022}. In our fiducial models, stellar populations with
$M_\star\sim10^{5-6}\,M_\odot$ can produce
${\rm DM}_{\rm HII}\gtrsim10^3\,{\rm pc\,cm^{-3}}$ for FRB progenitors that
remain embedded during the first several Myr of stellar evolution. These
columns arise on stellar-association scales and therefore need not be supplied
entirely by the immediate supernova or circumstellar environment.

As illustrated by Figure~\ref{fig:time_dependence}, this extreme-DM phase is
transient, persisting for only $\sim2$--$3\,{\rm Myr}$ after the onset of core
collapse. Thereafter, the declining ionizing photon production rate, continued
expansion of the H~II region, and mechanical evacuation by the growing
superbubble rapidly reduce the available electron column, reducing the DM at a rate consistent with $t^{-5}$.

This channel differs qualitatively from models in which the source-frame DM is
dominated by a young supernova remnant or dense circumstellar medium. In such
scenarios, producing ${\rm DM}\sim10^3\,{\rm pc\,cm^{-3}}$ generally requires
FRB activity to begin immediately after core collapse, while the ejecta or
circumstellar material remain compact and dense
\citep[e.g.,][]{piroImpactSupernovaRemnant2016, 2017ApJ...841...14M,
piroDispersionRotationMeasure2018, zhaoFRB190520BEmbedded2021}. By contrast,
the H~II-region contribution is sustained by the ionizing luminosity of the
surrounding stellar association and can remain large on Myr timescales. The FRB
source need not become active immediately after core collapse, provided that it
remains spatially associated with the ionized gas while the stellar population
continues to supply a substantial Lyman-continuum photon budget. As a corollary, we predict that non-repeating FRB sources in similar environments should also experience high DMs, if they have similar activation times to the repeater population.

The available H~II-region column is therefore controlled by the overlap of three
timescales: the core-collapse time of the neutron-star progenitor, the decline
of $\mathbb{N}_{\rm Lyc}(t)$, and the dynamical expansion of the ionized gas.
The most favorable systems are those in which the compact remnant forms early
and remains close to its birth association, allowing it to sample the ionized
gas before the local electron density dissipates. Later-forming
neutron stars, or sources that have drifted away from the association, may
still contribute substantially to the overall FRB population, but they
generally encounter smaller H~II-region dispersion measures in this channel. 

\subsection{Metallicity and the Preference for Dwarf Galaxies}

Metallicity strongly regulates the high-DM tail in our models.
At fixed stellar mass, lower-metallicity populations produce harder, more luminous ionizing spectra, while their lower dust content reduces the fraction of Lyman-continuum photons absorbed by grains. 
Both effects increase the ionized column available to embedded FRB sources.

Figure~\ref{fig:RHD_fiducial} shows this dependence on metallicity for fiducial associations with $M_\star=10^{5}$ and $10^{6}\,{\rm M_\odot}$. 
Milky-Way-like environments do not produce ${\rm DM}_{\rm HII}\gtrsim10^3\,{\rm pc\,cm^{-3}}$ through this channel, whereas LMC- and SMC-like systems can generate columns well above this threshold. 
This behavior provides a natural explanation for why the most extreme local-DM FRBs may preferentially occur in low-metallicity, star-forming dwarf galaxies
\citep[e.g.,][]{moroianuMilliarcsecondLocalizationAssociates2025, wangFastRadioBursts2026, loudasUnveilingOriginFast2025}.
Core-collapse supernovae and magnetar formation already trace recent star formation, but the H~II-region channel introduces an additional environmental selection: the largest local DMs are favored in systems that are both actively forming massive stars and sufficiently dust-poor to maintain large ionized columns.
The model therefore predicts that the extreme-DM tail should be overrepresented in low-metallicity dwarf hosts, even if the broader FRB population occupies a wider range of environments.

This metallicity dependence also distinguishes the H~II-region channel from a purely compact supernova-remnant or circumstellar explanation. Although dense CSM or young ejecta can enhance the DM around an individual source, the association-scale contribution depends on the collective ionizing photon budget and dust opacity of the surrounding stellar population. The model therefore provides a physical origin for the observed joint association between large source-frame DM, recent clustered star formation, and low-metallicity, dust-poor, environments~\citep{moroianuMilliarcsecondLocalizationAssociates2025, athukoralalage2026local}.

\subsection{Dependence on Key Parameters}
\label{sect:mass_and_dynamics}

Throughout this work we have adopted a set of fiducial parameters
appropriate to young, massive stellar clusters. In this section we examine
how the predicted dispersion-measure distribution responds to variations in a number of quantities left only loosely constrained by observations: the FRB
activity timescale $\tau_{\rm frb}$, the progenitor drift velocity
$v_{\rm drift}$, the association mass $M_\star$, and the upper limit on neutron star progenitor mass $M_{\rm NS,{\rm max}}$.

\subsubsection{Association Mass}

\begin{figure}[th]
    \centering
    \includegraphics[width=\linewidth]{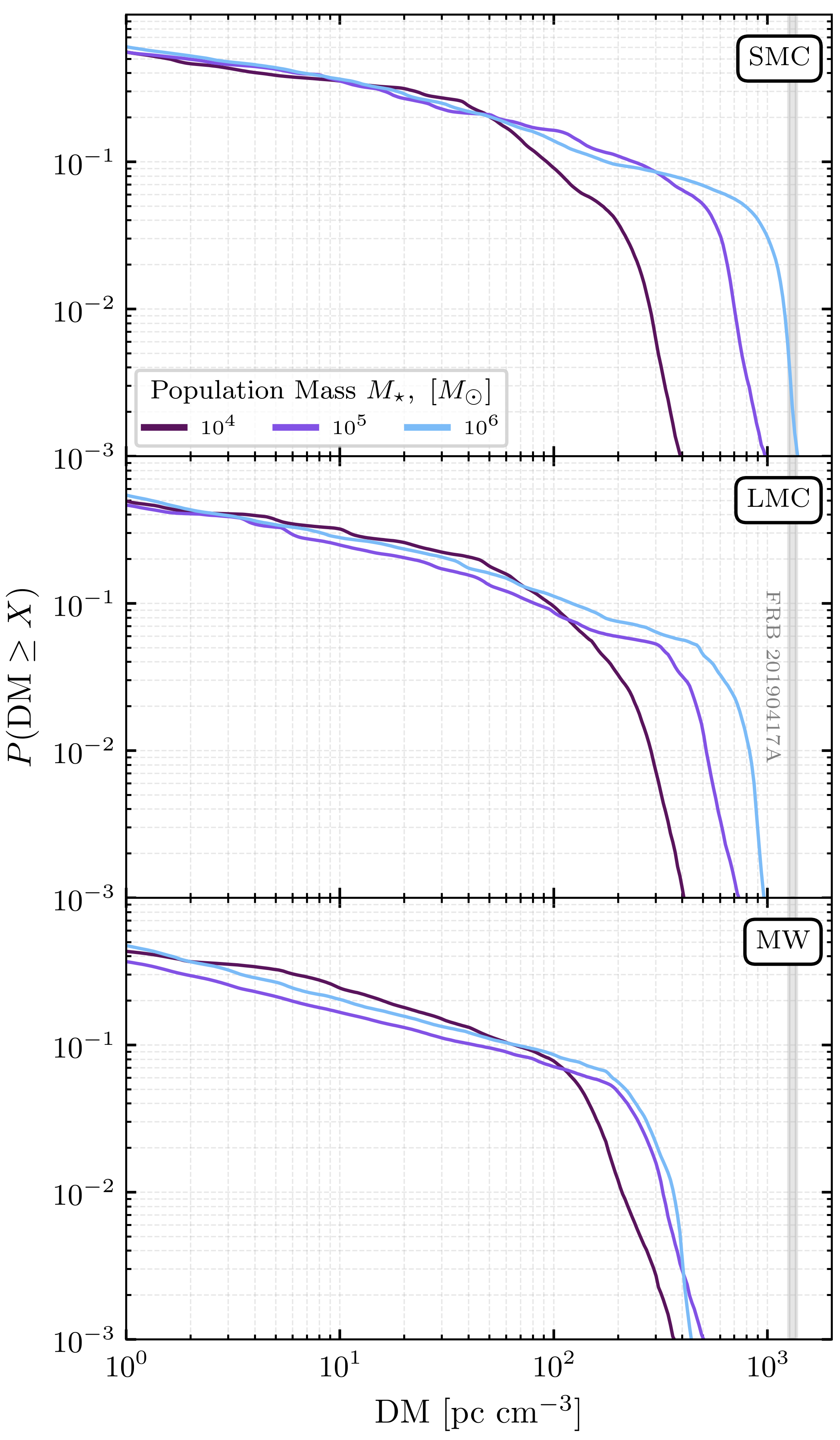}
    \caption{
    Dispersion measure survival functions, $P({\rm DM}\ge X)$, as a function of stellar population mass, $M_\star$, for each of the SMC-, LMC-, and Milky-Way-like fiducial associations (top to bottom). 
    In each panel, survival functions are plotted for 3 population masses: $10^4,\;10^5,\;$ and $10^6\;{\rm M_\odot}$. 
    All other parameters take their fiducial values.
    The rest-frame host DM of FRB20190417A is indicated for reference.}
    \label{fig:mass_comparison}
\end{figure}

The association mass $M_\star$ plays a central role in setting which formation scenarios can produce high-DM FRBs.
We repeat our RHD simulations for $M_\star=10^4$, $10^5$, and $10^6\,M_\odot$ across each of our three fiducial metallicities (SMC, LMC, and MW).
Naively, since $\mathbb{N}_{\rm Lyc}\propto M_\star$, one might expect higher-mass associations to produce systematically larger DMs; however, increasing $M_\star$ at fixed cloud surface density also decreases the ambient density, $n_{\rm H,0}\propto M_\star^{-1/2}$
(Equation~\ref{eq:n_H_scale}), which partially offsets the increased
photon budget. From Equation~\eqref{eq:dm_scaling}, the two effects combine
to give only a weak, $M_\star^{1/6}$ dependence of the early-time DM on
association mass. Consequently, while lower-mass associations do produce
systematically lower DMs, the effect is far weaker than the linear scaling
of $\mathbb{N}_{\rm Lyc}$ with $M_\star$ alone would suggest.

This mass dependence is most apparent at low metallicity, where
Figure~\ref{fig:mass_comparison} shows the high-DM population persisting,
but significantly reduced, down to $M_\star=10^4\,M_\odot$. At higher
metallicity, however, dust attenuation imposes an effective ceiling on the
achievable DM that increasing
$M_\star$ cannot overcome: because dust competes with recombination for the
same ionizing photon budget, and this competition strengthens with
increasing photon density, larger associations do not straightforwardly
produce proportionally larger dusty H\,II regions. As a result, association
mass alone is a comparatively weak lever for producing extreme DMs in
metal-rich environments, reinforcing that low metallicity is the more
important condition for this channel to reach
${\rm DM}\gtrsim10^3\,{\rm pc\,cm^{-3}}$.

\begin{figure}[t!]
    \includegraphics[width=\linewidth]{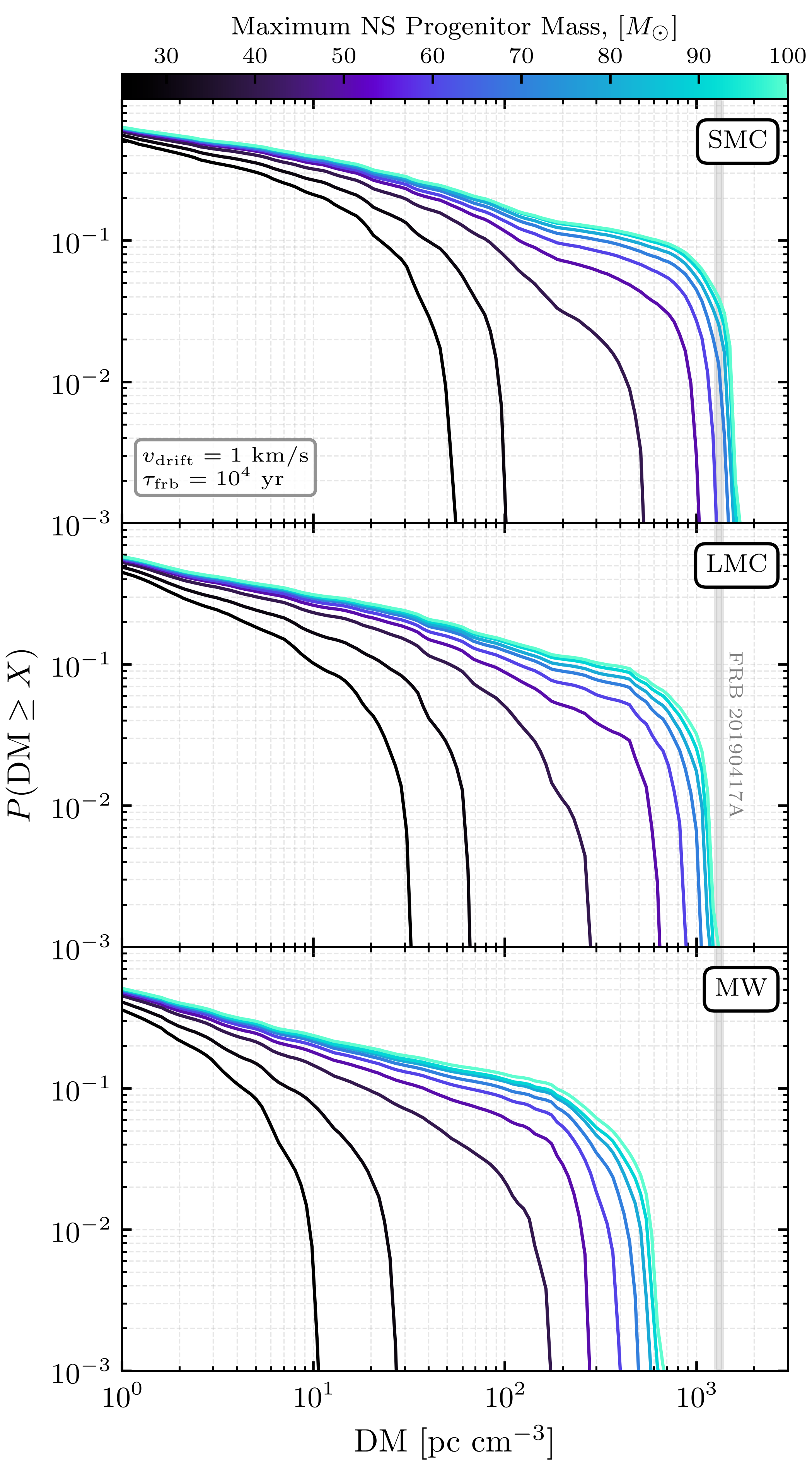}
    \caption{
    Same as figure~\ref{fig:mass_comparison} but for the maximum neutron star progenitor mass. Simulations were performed for $M_{\rm NS,max}$ ranging from $25$ to $100\;{\rm M_\odot}$ as indicated in the colorbar. All other parameters took their fiducial values, including $M_{\star} = 10^6\;{\rm M_\odot}$.}
    \label{fig:ns_mass_comparison}
\end{figure}

\subsubsection{Neutron Star Mass Threshold}

As discussed previously, the ultra-high dispersion measure phase of H~II region evolution is short lived ($\sim 3\;{\rm Myr}$) and is therefore only accessible to early forming neutron stars in the population. 
As a result, our model is quite sensitive to the formation epoch (and therefore progenitor ZAMS mass) of the first neutron stars.
While early theoretical calculations motivated a $25 \;M_\odot$ upper limit on the neutron star progenitor mass \citep{heger_how_2003}, more recent work has demonstrated that core-collapse outcomes depended quite sensitively on the pre-supernova state of the progenitor and are non-monotonic in ZAMS mass \citep{2016ApJ...821...38S, woosleyBirthFunctionBlack2020}. 
Our choice to adopt an effective neutron star progenitor mass range of $M\in [M_{\rm NS,min},M_{\rm NS,max}] = [9,60]\;{\rm M_\odot}$ is motivated by the existence of neutron stars with lower limits on progenitor mass exceeding $40\;M_\odot$ \citep{2006ApJ...636L..41M, 2009ApJ...707..844D, 2005ApJ...622L..49F}; however, the uncertain and non-monotonic mapping between ZAMS mass and compact-remnant outcome makes $M_{\rm NS,max}$ one of the least secure parameters in our population model. 
We therefore treat $M_{\rm NS, max}$ as an effective parameter controlling the earliest epoch of neutron-star formation, rather than as a literal boundary between neutron-star and black-hole formation, and explicitly explore the resulting DM distribution over a broad range of values.

Figure~\ref{fig:ns_mass_comparison} demonstrates the dependence of the FRB dispersion measure distribution on the maximum progenitor mass. 
Intuitively, lower cutoffs on neutron star formation delay the formation of the first neutron stars and therefore the first FRBs. 
These FRBs are then only able to sample the H~II region DM after the ionizing flux has already begun to decay and therefore fail to achieve the ultra-high DMs available to earlier forming neutron stars. 
For SMC-like environments, ${\rm DM} \gtrsim 1000\;{\rm pc/cm^3}$ require (assuming fiducial parameters) neutron star progenitors in excess of $\sim 50\;{\rm M_\odot}$, while ${\rm DM} \gtrsim 100\;{\rm pc/cm^3}$ require progenitor masses $\gtrsim 40\;{\rm M_\odot}$. 
The highest DM FRB, FRB 20190417A \citep{moroianuMilliarcsecondLocalizationAssociates2025}, would require $M_{\rm progenitor} \gtrsim 60\;{\rm M_\odot}$ to explain the entire DM excess attributed to the source.
If, as is most likely the case, some fraction of the DM is supplied by the diffuse ISM or by a host supernova remnant, then the constraint on $M_{\rm progenitor}$ need not be as stringent.

\begin{figure}[ht!]
    \centering
    \includegraphics[width=\linewidth]{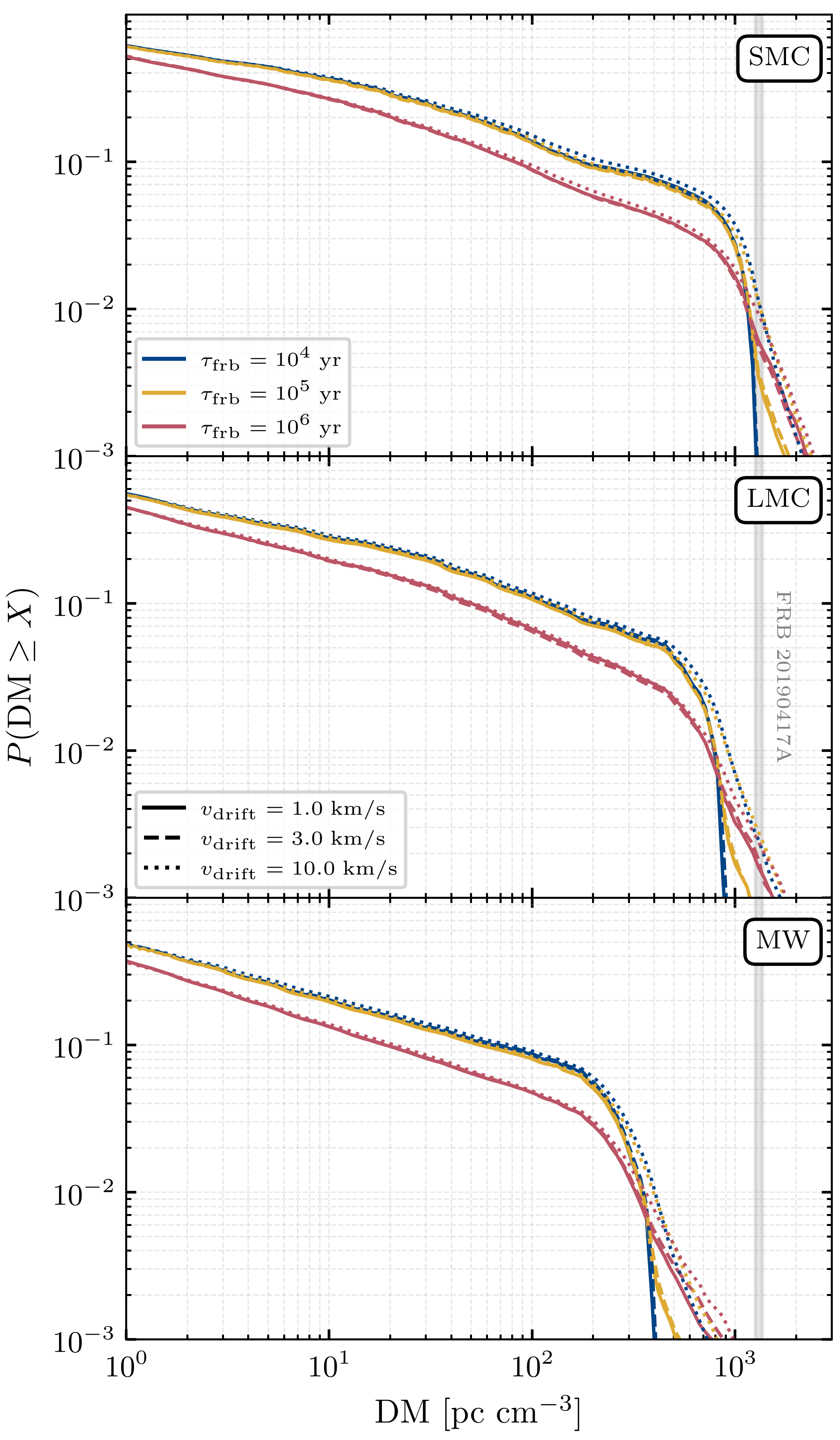}
    \caption{
    Dispersion measure survival functions, $P({\rm DM}\ge X)$, as a function of the FRB activity timescale, $\tau_{\rm frb}$, and the main-sequence drift velocity, $v_{\rm drift}$, for each of the SMC-, LMC-, and Milky-Way-like fiducial associations (top to bottom). All simulations were performed for a $10^6\;{\rm M_\odot}$ association. Each panel shows $P({\rm DM} > X| \tau_{\rm frb}, v_{\rm drift})$ for $\tau_{\rm frb} \in \{10^4,10^5,10^6\}\;{\rm yr}$ and $v_{\rm drift} \in \{1, 3, 10\}\;{\rm km/s}$. For $\tau \ll 10^6\;{\rm yr}$, results are largely insensitive to the parameter. Variations in $v_{\rm drift}$ are most prominent in broadening the high DM tail of the distribution.}
    \label{fig:velocity_comparison}
\end{figure}

\subsubsection{FRB Activation Time and Drift Velocity}

The preceding section considered the earliest time at which an FRB-capable neutron star can form, as set by the progenitor mass and its main-sequence lifetime. 
Core collapse, however, does not uniquely determine either the time or radius at which the source samples the H~II region. 
After the progenitor is born, an additional delay may separate core collapse from the onset of FRB activity, while stellar motions before collapse and natal kicks afterward can spatially displace the source from the dense ionized gas. The observed DM therefore depends not only on when neutron stars form, but also on when they become active and whether they remain embedded in their natal environment. The two most important quantities that set that
activation time and position, $\tau_{\rm frb}$ and $v_{\rm drift}$, are among the least observationally constrained parameters in the model.

We test the sensitivity of our results to these choices by independently
varying $\tau_{\rm frb}\in\{10^4,10^5,10^6\}\,{\rm yr}$ and
$v_{\rm drift}\in\{1,3,10\}\,{\rm km\,s^{-1}}$  for each of our
SMC-, LMC-, and Milky-Way-like fiducial associations.
Figure~\ref{fig:velocity_comparison} shows the resulting DM distributions.
We find that $\tau_{\rm frb}$ has little effect on the distribution as long
as it remains short compared to the core-collapse timescale of the
association ($\sim$few Myr). Only once
$\tau_{\rm frb}$ approaches $10^6\,{\rm yr}$, comparable to the timescale
over which the H\,II region and superbubble themselves evolve, does the delay time have a significant impact on the dynamics. This is primarily driven by dissociation of FRB progenitors from the natal association due to their random kicks following core collapse. For progenitors with long delay times, the high kick velocity can easily remove many of the progenitors from the association entirely.

By contrast, $v_{\rm drift}$ has only a modest effect on the overall DM distribution.
Its influence is most apparent at low DM, where the sources are typically older, lower-mass progenitors with longer main-sequence lifetimes and therefore more time to drift from their birth sites.
Increasing $v_{\rm drift}$ broadens their spatial distribution and allows a larger fraction to move partially or entirely out of the natal cloud, reducing their H~II-region contribution to the DM.
The high-DM population is much less sensitive to $v_{\rm drift}$, although modest displacements can slightly broaden its peak by placing some massive progenitors along longer sightlines through the ionized gas.

\section{Discussion}
\label{sect:disc}

\subsection{Dispersion Across Multiple Environmental Scales}
\label{sect:multichannel_dm}

The H~II-region contribution considered in this work represents one component
of a broader hierarchy of electron reservoirs surrounding an FRB source. The
source-frame host dispersion measure may be written schematically as
\begin{equation}
    {\rm DM}_{\rm host}
    =
    {\rm DM}_{\rm eng}
    +
    {\rm DM}_{\rm SNR/CSM}
    +
    {\rm DM}_{\rm HII}
    +
    {\rm DM}_{\rm ISM},
    \label{eq:DM_comps}
\end{equation}
where the individual terms arise on progressively larger spatial and temporal
scales. These components can coexist, and the boundaries between them are not
necessarily sharp. Their relative importance depends on the FRB emission
mechanism, progenitor mass-loss history, explosion environment, activation
delay, and subsequent motion of the compact remnant. Representative
contributions are as follows:

\begin{itemize}

    \item \textbf{Engine-scale and Magnetospheric Contributions
    ($r\sim10^6-10^{\rm 10}\;{\rm cm}$).}
    Plasma associated with the magnetosphere, magnetar wind, or a compact
    nebula can contribute on sub-parsec scales. The magnitude of this term is
    highly model dependent because relativistic motion, pair multiplicity,
    magnetic-field geometry, and the location of the radio-emission region can
    all modify the effective dispersive column. Compact nebulae may also be
    associated with persistent synchrotron emission and large rotation
    measures, although neither observable uniquely determines their DM
    contribution.

    \item \textbf{Supernova-remnant Contributions
    ($10^{14}-10^{18}\;{\rm cm}$).}
    The ejecta, shocked circumstellar material, and swept-up forward-shock
    shell produced at core collapse can provide a substantial ionized column
    during the early post-explosion phase
    \citep[e.g.,][]{piroImpactSupernovaRemnant2016,
    piroDispersionRotationMeasure2018,
    zhaoFRB190520BEmbedded2021}. Reaching the upper end of this range generally
    requires rapid FRB activation, $\tau_{\rm FRB}\lesssim10^4\,{\rm yr}$,
    together with dense ejecta or a dense circumstellar environment. The
    contribution subsequently declines as the remnant expands and its column
    is diluted. The detailed evolution depends on the ejecta mass and energy,
    progenitor wind profile, external density, ionization state, and continued
    energy injection from the central compact object. At sufficiently long
    delays, natal kicks may also displace the source from the highest-column
    regions of the remnant.

    \item \textbf{Stellar-population contribution
    ($10^{18}-10^{21}\;{\rm cm}$).}
    The natal H~II region is maintained by the collective ionizing luminosity of
    the surrounding stellar association and therefore evolves on
    association-scale, Myr timescales. Its contribution depends on the
    association mass, ambient gas density, metallicity, dust opacity, and the
    position of the FRB relative to the ionized gas. The calculations presented
    here show that massive, low-metallicity associations can reach
    ${\rm DM}_{\rm HII}\gtrsim10^3\,{\rm pc\,cm^{-3}}$, even after the
    immediate supernova remnant has become too dilute to dominate.
    Progenitor drift, natal kicks, cloud breakout, and mechanical evacuation by
    a superbubble reduce this contribution by separating the source from the
    dense ionized gas or redistributing that gas into expanding shells.

    \item \textbf{Diffuse host-galaxy contribution
    ($\gtrsim 10^{21}\;{\rm cm}$).}
    Ionized gas in the galactic disk, diffuse ISM, halo, outflows, and unrelated
    star-forming structures contributes over scales of hundreds of parsecs to
    tens of kiloparsecs. Typical host contributions are of order several tens to $\sim100\,{\rm pc\,cm^{-3}}$ from observations~\citep{leung2025stellar} and galaxy-scale hydrodynamical simulations~\citep{orr_objects_2024} although individual
    sightlines can extend to substantially larger values. It depends on
    the host morphology, inclination~\citep{cassanelli2024fast}, star-formation rate~\citep{li2025calibrating}, and the location of the FRB within the galaxy (see hyper-local sources such as~\citep{2025ApJ...989L..48C}) and should generally
    vary much more slowly than the compact source components.
\end{itemize}

These reservoirs need not act independently. A young neutron star may
simultaneously be embedded in a compact nebula, expanding supernova ejecta, a
natal H~II region, and the diffuse host ISM. The observed source-frame DM can
therefore contain a slowly varying association- or galaxy-scale baseline
together with a more rapidly evolving compact contribution. The dominant term
is also expected to change with source age: engine-scale plasma and dense
ejecta are most relevant shortly after core collapse, the natal H~II region can
remain important over the subsequent Myr-scale core-collapse epoch, and the
diffuse host ISM may dominate once the remnant has expanded and the source has
separated from its birth environment.

The recent observation of secular DM variations on year-to-multi-year timescales in repeating FRBs \citep{cookDiscovery30Repeating2026, zhaoFRB190520BEmbedded2021} indicates that at least part of the dispersing column arises in a compact, dynamically evolving environment. 
The variability timescale places a model-independent causal upper limit of order $R\lesssim ct_{\rm var}\sim10^{18}\,{\rm cm}$ for $t_{\rm var}\sim1\,{\rm yr}$, consistent with plasma associated with an evolving supernova remnant or other near-source structure.
In principle, improved models of the individual DM reservoirs, together with a better understanding of the FRB emission mechanism and source evolution, may allow the compact, time-variable contribution to be separated from the more slowly varying H~II-region and host-galaxy components. At present, however, uncertainties in the source geometry, ionization state, plasma dynamics, and progenitor history prevent a unique decomposition.

\subsection{Implications for Cosmological Inference} 

A major motivation for FRB population studies is their use as probes of the
ionized intergalactic medium (IGM), and hence of the cosmic baryon distribution. For a localized FRB, the observed
dispersion measure can be written schematically as
\begin{equation}
    {\rm DM}_{\rm obs}
    =
    {\rm DM}_{\rm MW}
    +
    {\rm DM}_{\rm IGM}
    +
    \frac{{\rm DM}_{\rm host}}{1+z},
    \label{eq:dm_cosmology_decomposition}
\end{equation}
where ${\rm DM}_{\rm MW}$ includes the Milky Way disk and halo,
${\rm DM}_{\rm IGM}$ is the intergalactic contribution, and
${\rm DM}_{\rm host}$ includes both the diffuse host-galaxy ISM and plasma
associated with the immediate source environment. This expression neglects
additional intervening halos or galaxy clusters along the line of sight, but
captures the basic degeneracy relevant for cosmological applications.

The principal difficulty is therefore not simply estimating the diffuse
host-galaxy ISM contribution, but marginalizing over a set of unresolved
electron reservoirs whose relative importance depends on source age,
progenitor history, stellar environment, and viewing geometry. A model that
equates ${\rm DM}_{\rm host}$ with the diffuse ISM alone may substantially
underestimate both the width and the high-DM tail of the host contribution.
The results presented here identify young, massive, low-metallicity stellar
associations as one physically motivated source of that tail and therefore
suggest that host-DM priors should account for the local star-forming
environment of the FRB.

\subsection{Observational Tests}

If the model presented in this work is indeed correct and extreme-DM FRBs arise from sources embedded in H~II regions, then fairly stringent requirements can be placed on local stellar population. 
Specifically, the large electron columns in our models occur only while the FRB remains embedded in dense ionized gas and the surrounding stellar population retains a substantial Lyman-continuum photon budget.
The relevant counterpart should therefore exhibit both nebular emission from the ionized gas and continuum emission from a stellar population with an age of at most several Myr. 
Both of these emission components may be observable and could serve as direct tests of the model in nearby hosts.

The most direct tracer of the dispersing gas is recombination-line emission.
For ionization-bounded gas in approximate case-B equilibrium, the intrinsic H$\alpha$ luminosity is
\begin{equation}
    L_{\rm H\alpha}^{\rm int}
    =
    h\nu_{\rm H\alpha}
    \frac{\alpha_{\rm H\alpha}^{\rm eff}}{\alpha_B}
    \mathbb{N}_{\rm rec},
    \label{eq:halpha_luminosity_general}
\end{equation}
where $\mathbb{N}_{\rm rec}$ is the rate of hydrogen recombinations within the nebula.
In the uniform-density model described in Section~\ref{sect:modeling},
\begin{equation}
    \mathbb{N}_{\rm rec}
    =
    y_{\rm dust}^{3}\mathbb{N}_{\rm Lyc},
\end{equation}
because dust reduces the ionized volume, and hence the total recombination
rate, by a factor $y_{\rm dust}^{3}$ at fixed ambient density. At
$T\simeq10^{4}\,{\rm K}$,
\begin{equation}
    L_{\rm H\alpha}^{\rm int}
    \simeq
    1.4\times10^{40}\,
    y_{\rm dust}^{3}
    M_{\star,6}Q_{\rm Lyc,46}
    \ {\rm erg\,s^{-1}}.
    \label{eq:halpha_luminosity_dusty}
\end{equation}
Depending on the diffuse extinction properties of the host galaxy, such a large ${\rm H}\alpha$ luminosity may be observable out to significant redshift, far exceeding the scales on which the local region of the FRB remains resolvable.

H$\alpha$ detection establishes the presence of photoionized gas, but does
not by itself uniquely determine the age or stellar mass of the population
responsible for the ionization. A stronger test would combine spatially
resolved or astrometrically localized recombination-line measurements with
UV--optical photometry. For the fiducial
$M_\star=10^6\,{\rm M_\odot}$ associations, the
{\tt pyStarburst99} spectra predict a luminous near-UV counterpart during
the high-DM phase, with
\begin{equation}
    \nu L_\nu \sim 10^{41}\ {\rm erg\,s^{-1}}
\end{equation}
near the pivot wavelengths of the HST/WFC3 F225W and F275W filters.
The near-UV continuum is dominated by massive and intermediate-mass stars
and therefore provides a complementary tracer of recent star formation,
while the H$\alpha$ luminosity responds more directly to the shorter-lived
ionizing population.

The characteristic H~II-region radii in our models are of order
$\sim100\;{\rm pc}$, and in most cases the nebula will therefore be unresolved
or only marginally resolved. Full morphological resolution is not required,
however, provided that the FRB position can be localized to a sufficiently
small portion of the surrounding star-forming complex. The relevant
observational limitation is the degree to which nebular and stellar emission
associated with the FRB can be separated from neighboring H~II regions,
diffuse ionized gas, and older stellar populations.

For nearby sources, progressively finer physical resolution provides a
cleaner measurement of the local H$\alpha$ luminosity, UV continuum, and
stellar-population age. Even when the $\sim100\;{\rm pc}$ H~II region is not
resolved into multiple resolution elements, precise astrometric coincidence
between the FRB and a compact, UV-bright, recombination-line-emitting source
would provide strong support for the model. By contrast, an unresolved
measurement dominated by a larger star-forming complex is more difficult to
interpret, because contamination can bias the inferred age, extinction, and
ionizing luminosity away from those of the immediate FRB environment.

\subsection{Limitations and Assumptions}

The model developed in this work is intentionally idealized, with the goal of assessing whether natal H~II regions can plausibly contribute
$\gtrsim 10^3\,{\rm pc\,cm^{-3}}$ to FRB source-frame dispersion measures.
Although it includes the principal physical ingredients, several effects are omitted because they are either poorly constrained or would add complexity beyond the scope of this toy model.
Here we summarize the most important assumptions and discuss their qualitative impact on the results.

\paragraph{Dust and Radiative Transfer} 
Dust absorption of ionizing photons plays a central role in setting the metallicity dependence of the H~II-region dispersion measures in our model.
At fixed stellar mass and gas density, dust competes with neutral hydrogen for Lyman-continuum photons, reducing the ionized volume and suppressing the resulting electron column. In the absence of this effect, the contrast between low- and high-metallicity environments would be substantially weaker, and massive associations in metal-rich hosts would more readily produce ${\rm DM}_{\rm HII}\sim10^3\,{\rm pc\,cm^{-3}}$.
This introduces an important systematic uncertainty: the quantity most relevant for our model is not the galaxy-integrated dust mass or optical attenuation, but the effective Lyman-continuum dust absorption opacity per hydrogen atom in the local FRB birth environment. This opacity is rarely measured directly and need not be well represented by Milky-Way, LMC, or SMC dust templates. \citep[see e.g.][]{2018ARA&A..56..673G, 1994ApJ...429..582C}.
Galaxy-integrated dust masses can be inferred from broad-band spectral-energy distributions and infrared emission, but such estimates depend on assumptions about dust temperature, emissivity, composition, and grain-size distribution \citep[see e.g.][]{draine2011physics, 2005pcim.book.....T,2001AJ....122.1788I, 2022ApJ...931...14L}.
Moreover, these observables constrain the total dust content only indirectly related to the quantity most relevant here: the effective Lyman-continuum absorption cross section per hydrogen atom in the local H~II-region environment.
Because this opacity is rarely measured directly, especially in low-metallicity dwarf galaxies, the dust prescription remains one of the dominant uncertainties in predicting both the H~II-region DM and the associated recombination-line luminosity.

\paragraph{Stellar Physics in Dwarf Galaxies} 

A key feature of the H~II-region channel is the timing overlap between three
clocks: the decline of the stellar Lyman-continuum luminosity, the onset of
core collapse for neutron-star progenitors, and the activation time of the FRB
source. The highest H~II-region DMs occur only when the ionizing photon budget
remains large at the time when FRB-capable compact remnants are present. As a
result, uncertainties in the amplitude and duration of $N_{\rm Lyc}(t)$ map
directly into uncertainties in the predicted high-DM tail.
Because our results are sensitive systematic effects in $N_{\rm Lyc}$, we opted to use {\tt pySTARBURST99}, which incorporates many improvements over older models \citep[e.g.,][]{1999ApJS..123....3L}, including updated low-metallicity evolutionary tracks, rotation, very massive stars, and a new grid of {\tt FASTWIND} stellar-atmosphere models \citep{2025ApJS..280....5H}.
Nevertheless, the relevant massive-star physics remains uncertain,
especially in the low-metallicity regime \citep[e.g.,][]{2024ARA&A..62...21M, Eldridge_2022} and when accounting for the effects of binarity \citep{2017PASA...34...58E, 2009MNRAS.400.1019E}.
Exemplifying our uncertainty in this regard, the commonly used line-driven wind
prescription of \citet{2001A&A...369..574V} has recently been challenged by
observations of extremely metal-poor O stars, which indicate substantially
weaker winds (and stronger ionizing flux) than predicted by the standard prescription \citep{2024ApJ...974...85T}.

In addition to uncertainties concerning the ionizing flux of low-metallicity populations, our assumption of a \citet{2001MNRAS.322..231K} IMF introduces another potential systematic \citep[e.g.,][]{2017MNRAS.468..319E, 2013ApJ...771...29G, 2019ARA&A..57..375S, 2012ApJ...747...72D}.
It has been broadly conjectured that the IMF of low-metallicity dwarf galaxies may be top-heavy \citep[e.g.,][]{2012ApJ...747...72D, 2012MNRAS.422.2246M}; however, this remains debated in the literature \citep{2022MNRAS.509.1959S}.
If the IMF of low-metallicity dwarfs is indeed top-heavy relative to that of the Milky Way, this will significantly increase the frequency of high-dispersion FRBs from H~II regions in those systems as the most massive neutron star progenitors which form FRBs are also those which produce the highest dispersion.
Nonetheless, we do not consider variations in the IMF in this work.

\paragraph{Stellar Feedback}

Throughout this work, we have made the assumption that mechanical feedback driving H~II region dissipation is primarily provided by supernovae, following the typical energy-driven superbubble treatment \citep{1975ApJ...200L.107C, 1977ApJ...218..377W}.
It is, however, well established that stellar feedback is considerably more complex than this simple picture would indicate, and that our implementation is only an approximation.
Direct and reprocessed radiation pressure \citep{2009ApJ...703.1352K, 2011ApJ...732..100D}, stellar winds \citep{2020A&A...639A...2P, 2016MNRAS.456..710F}, and geometry \citep{2025MNRAS.540.1124L} all play important roles that are not accounted for in this work.
Qualitatively, we do not expect the details of cloud dispersion to impact the ability of H~II regions to produce high early-time DMs; however, differences in the treatment of these details may substantially change the late-time behavior of the model.
In low-metallicity environments, our exclusion of wind-driven feedback, specifically, is better justified, as massive stars are expected to have lower mass-loss rates in this regime \citep{2012ApJ...751L..34V}; this argument does not extend to radiation pressure or to the geometric idealization discussed above.
Future refinements of this model may be able to significantly improve the treatment of cloud dispersal (see, e.g., \citealt{2017MNRAS.470.4453R, 2026arXiv260527517T}).

\paragraph{FRB Emission Mechanisms} 
A final systematic uncertainty is the unknown physical mechanism that produces FRBs. 
Magnetar models provide a natural connection between FRBs, massive-star formation, and core-collapse environments, but the emission mechanism, activation delay, and active lifetime remain uncertain \citep[e.g.,][]{2017ApJ...841...14M, 2019MNRAS.485.4091M, 2020MNRAS.498.1397L, beloborodovBlastWavesMagnetar2020}. 
Our model is therefore intentionally agnostic about the detailed engine. It requires only that some FRB-capable compact remnants form in young stellar associations and become active while still spatially associated with their natal ionized gas. 
The predicted H~II-region DM is only weakly sensitive to short activation delays. Delays of $10^3$--$10^5\,{\rm yr}$ are small compared with the Myr-scale evolution of the H~II region, and therefore produce similar local columns.
Delays approaching $\sim10^6\,{\rm yr}$, however, can matter: the H~II region may have expanded, the electron density may have declined, and the remnant may have drifted away from the highest-column gas. 

The larger uncertainty is not the precise prompt delay, but the mapping from massive-star progenitors to FRB-active remnants. 
If FRB production depends on progenitor mass, metallicity, binarity, rotation, or explosion outcome, then the FRB formation time will be weighted toward different parts of the declining $N_{\rm Lyc}(t)$ curve.
This can change the fraction of sources that sample large H~II-region columns.
For this reason, our results should be interpreted as predictions for the environmental DM available to young FRB-capable remnants, rather than as a theory for the absolute formation rate or activation history of FRB engines, for which a detailed understanding of FRB progenitor physics is necessary.

\section{Conclusions}
\label{sect:conclusion}

This work demonstrates that extreme source-frame dispersion measures in FRBs can be produced by natal H~II regions surrounding young massive stellar associations.
Rather than requiring the full local column to arise from a compact supernova remnant or dense circumstellar medium, the model developed here shows that association-scale ionized gas can provide ${\rm DM}_{\rm HII}\sim10^3\,{\rm pc\,cm^{-3}}$ for FRB progenitors that form early and remain spatially associated with their birth environment.

We constructed a semi-analytic model for FRB progenitors embedded in cluster-forming molecular clouds. The model connects the stellar birth mass to the ionizing photon budget, maps the IMF onto a time-dependent neutron-star birth rate, accounts phenomenologically for FRB activation after core collapse, and follows the spatial decorrelation of progenitors from their natal gas. We then coupled this progenitor model to a dusty, expanding H~II region, deriving the electron columns expected for sources embedded at different locations and activation times. This analytic framework identifies the essential controls on the local DM: the stellar mass of the association, the gas density, the dust absorption of Lyman-continuum photons, the time evolution of the ionizing luminosity, and the degree of progenitor drift.

We supplemented this framework with idealized one-dimensional
radiation-hydrodynamic simulations that follow the response of the gas to the
time-dependent ionizing luminosity of an evolving stellar population. These
calculations include photoionization heating, pressure-driven expansion, Lyman-continuum dust absorption, and stellar spectra from
\texttt{pySTARBURST99}. They confirm the basic analytic picture while capturing the rapid decline of $N_{\rm Lyc}(t)$, the expansion of the ionized region, and the changing electron column encountered by FRB progenitors during the core-collapse epoch.

Our principal finding is that massive, young, low-metallicity stellar
associations provide a natural environment for producing the extreme-DM tail of the FRB population. In fiducial models with
$M_\star\sim10^6\,M_\odot$, the IZw18-like and SMC-like cases can maintain
H~II-region columns of order ${\rm DM}_{\rm HII}\sim10^3\,{\rm pc\,cm^{-3}}$
during the epoch when the first neutron-star progenitors undergo core collapse.
The corresponding Milky-Way-like models are more strongly suppressed because
dust absorbs a larger fraction of the Lyman-continuum photon budget before it
can ionize hydrogen. The model therefore predicts that the largest local DMs
should preferentially occur in environments that are both actively forming
massive stars and sufficiently dust-poor to sustain extended ionized gas.

This scenario provides a natural interpretation for low-redshift, high-DM FRBs in star-forming dwarf galaxies. It does not require all FRBs, or even all young magnetars, to reside in massive H~II regions. Instead, it identifies a specific environmental channel: compact remnants born early in massive, low-metallicity associations can sample large ionized columns for a substantial fraction of the massive-star core-collapse epoch. In this sense, the H~II-region contribution is complementary to supernova-remnant and circumstellar-medium models. A given FRB may receive dispersion from several local reservoirs, but the association-scale H~II component can by itself reach the values required for the most extreme systems.

Several important uncertainties remain. Unresolved
multi-dimensional cloud structure, low-metallicity dust physics, massive-star
evolution, binarity, IMF variations, and the mapping from neutron-star birth to FRB activity can all modify the quantitative rate and normalization of this
channel.
Nevertheless, the central conclusion is robust: young massive stellar
associations in low-metallicity, dust-poor environments can generate local
electron columns large enough to explain the most extreme source-frame
dispersion measures observed in FRBs. The highest-DM FRBs may therefore be not only probes of compact remnant environments, but also signposts of recent
clustered massive-star formation in dwarf galaxies.

\begin{acknowledgments}
\end{acknowledgments}
This material is based upon work supported by the National Science Foundation Graduate Research Fellowship Program under Grant No. DGE 2146752. Any opinions, findings, and conclusions or recommendations expressed in this material are those of the author(s) and do not necessarily reflect the views of the National Science Foundation. 
C. L. acknowledges support from the Miller Institute for Basic Research at UC Berkeley.
We thank Chris McKee for helpful discussions.

\begin{contribution} 
ECD was responsible for conceptualization, methodology, software development, formal analysis, visualization, and writing of the original manuscript. CL contributed to conceptualization, scientific interpretation, supervision, and manuscript review and editing. WL contributed through scientific discussion and interpretation of the results.



\end{contribution}

%
\software{
This work utilized {\tt NumPy} \citep{oliphant2006guide},
{\tt SciPy} \citep{2020SciPy-NMeth},
and {\tt AstroPy} \citep{2022ApJ...935..167A} for scientific and astronomical computations.
{\tt Numba} \citep{lam2015numba} was use to accelerate performance-critical elements of the analysis and simulation pipelines.
Figures were generated using {\tt Matplotlib} \citep{Hunter:2007} and the perceptually uniform colormaps of {\tt CMasher} \citep{2020JOSS....5.2004V}.
We acknowledge the use of {\tt Claude Code} to assist in software analysis, debugging and code documentation.
}



\appendix

\section{Numerical Method for the Idealized H~II-Region Simulations}
\label{app:numerical_method}

In this appendix we describe the numerical method used for the idealized
one-dimensional radiation-hydrodynamic simulations introduced in
Section~\ref{sect:numerical}. The goal of these simulations is not to model the
full multi-dimensional structure of star-forming regions, but rather to
capture the leading-order response of an H~II region to a time-dependent
ionizing photon production rate. This provides a controlled way to estimate
the electron columns encountered by FRB progenitors embedded in evolving
stellar associations.

The calculation is organized as an operator-split radiation-hydrodynamic
scheme. At the start of each timestep, the radiation step updates the
ionization structure using the current gas density profile, the instantaneous
ionizing photon production rate of the stellar population, and an effective
dust absorption opacity for Lyman-continuum photons. The resulting ionized
fraction profile determines the gas pressure entered into the equation of
state, which the hydrodynamic step then uses to evolve the spherically
symmetric gas density and radial velocity for the remainder of the timestep.

\subsection{Gas Dynamics}

We solve the spherically symmetric isothermal Euler equations for the gas
density $\rho(r,t)$ and radial velocity $v(r,t)$. In conservative form,
these are
\begin{align}
    \frac{\partial \rho}{\partial t}
    + \frac{1}{r^2}\frac{\partial}{\partial r}
    \left(r^2 \rho v\right)
    &= 0,
    \label{eq:hydro_continuity}
    \\
    \frac{\partial (\rho v)}{\partial t}
    + \frac{1}{r^2}\frac{\partial}{\partial r}
    \left[r^2(\rho v^2 + P)\right]
    &= \frac{2P}{r}.
    \label{eq:hydro_momentum}
\end{align}
The source term on the right-hand side of Equation~\eqref{eq:hydro_momentum}
is the geometric term that appears when the radial pressure gradient is
written in spherical finite-volume form, $\partial P/\partial r =
r^{-2}\partial(r^2 P)/\partial r - 2P/r$.

The equations are evolved on a one-dimensional spherical grid using a
first-order Godunov method with HLL fluxes. This low-order scheme is
sufficient for the present purpose because we are interested primarily in the
large-scale density and ionization structure that determine the integrated
dispersion measure, rather than the detailed structure of the ionization-front
transition. Integrating Equation~\eqref{eq:hydro_momentum} over a spherical
shell of volume $V_i$ bounded by faces of area $A_{i-1/2}$ and $A_{i+1/2}$
gives the geometric source as an exact area integral.
Approximating $P$ as piecewise constant within each cell evaluates this integral exactly, so the discrete
update takes the well-balanced form
\begin{align}
    \Delta \rho_i &= -\frac{\Delta t}{V_i}
        \Bigl[A_{i+1/2} F_{\rho,i+1/2} - A_{i-1/2} F_{\rho,i-1/2}\Bigr],
    \\
    \Delta (\rho v)_i &= -\frac{\Delta t}{V_i}
        \Bigl(\bigl[A_{i+1/2} F_{{\rm mom},i+1/2} - A_{i-1/2} F_{{\rm mom},i-1/2}\bigr]
        - P_i\,(A_{i+1/2} - A_{i-1/2})\Bigr),
    \label{eq:wellbalanced_update}
\end{align}
where $F_\rho = \rho v$ and $F_{\rm mom} = \rho v^2 + P$ are evaluated at each
face with the HLL Riemann solver. 

An artificial viscosity is added to damp post-shock oscillations. Following
the von Neumann--Richtmyer \citep{vonneumannMethodNumericalCalculation1950} prescription, a compressive-only bulk viscosity
pressure
\begin{equation}
    q_i = C_{\rm AV}\,\rho_i\,\bigl[\min(\delta v_i, 0)\bigr]^2,
    \qquad
    \delta v_i = \tfrac{1}{2}(v_{i+1}-v_{i-1})
\end{equation}
is added to the thermal pressure, $P_{{\rm eff},i} = P_i + q_i$, and used in
place of $P_i$ in the HLL flux and wave-speed estimates (but not in the
geometric source term of Equation~\eqref{eq:wellbalanced_update}, since the
artificial viscosity is a numerical dissipation term rather than a real
radial body force). The coefficient $C_{\rm AV}$ is an order-unity tunable
parameter; $q_i=0$ identically in expanding flow. In all simulations, $C_{\rm AV} = 20$ was used; however, only minimal differences were found for $1 \lesssim C_{\rm AV} \lesssim 30$.

The gas pressure is specified using a two-temperature equation of state.
Each cell is assigned an ionized fraction $x$, where $x=1$ corresponds to
fully ionized hydrogen and $x=0$ corresponds to neutral gas, and the ionized
and neutral phases are assigned separate mean molecular weights $\mu_{\rm
II}$ and $\mu_{\rm cloud}$ (in units of $m_{\rm H}$) set by the assumed
hydrogen, helium, and metal mass fractions of the gas, with only hydrogen
assumed to ionize. The pressure is then
\begin{equation}
    P =
    \frac{\rho k_{\rm B}}{\mu_{\rm I}m_{\rm H}}
    \left[
        \frac{\mu_{\rm I}}{\mu_{\rm II}}xT_{\rm II}
        +(1-x)T_{\rm I}
    \right],
    \label{eq:two_temperature_pressure}
\end{equation}
where $T_{\rm I}$ and $T_{\rm II}$ are the neutral- and ionized-gas
temperatures, respectively, and $\mu_{\rm I}$ and $\mu_{\rm II}$ are the
corresponding mean molecular weights.

Because Equation~\eqref{eq:two_temperature_pressure} provides no explicit
source of interior support beyond thermal pressure, a purely thermal balance
can, once the photoionized interior's density, and hence pressure, has
diluted through the general expansion, allow the higher-pressure exterior gas
to drive a spurious recollapse of already-ionized gas at late times, a regime
this isothermal treatment has no physics to prevent on its own. We suppress
this artifact with a kinematic constraint: after the conservative update,
cells with ionized fraction above a threshold $x > 0.5$ are not
permitted to have inward radial velocity. This clamp is applied purely as a
floor on $v$ after the flux update, so it does not alter density or pressure
and hence does not affect the CFL-limited timestep; its cost is a small,
local violation of momentum conservation in cells where it is triggered. This clamp was validated against the known Spitzer solution \citep{1978ppim.book.....S} to ensure no spurious behaviors arose.

\subsection{Operator-Split Ionization Update}

The radiation field is coupled to the gas using an operator-split ionization
update. At the beginning of each hydrodynamic timestep, we compute the
ionization structure from the current gas density profile and the instantaneous
ionizing photon production rate $\mathbb{N}_{\rm Lyc}(t)$. We adopt the
on-the-spot approximation and assume a sharp ionization front, so that
recombinations to the ground state are locally reprocessed and the ionized
region is determined by case-B ionization balance. This operator-split
treatment is justified because the recombination timescale,
$t_{\rm rec}=(\alpha_B n_{\rm H})^{-1}\sim10^5\,{\rm yr}$ for the densities of
interest, is short compared with the dynamical timescale of the H~II region,
$t_{\rm dyn}\sim R/c_s\sim10^7\,{\rm yr}$, so the ionization state equilibrates
essentially instantaneously relative to the gas motion.

Formally, the budget of available ionizing photons at radius $r$ from the
origin is required to obey
\begin{equation}
    \frac{dN_{\rm Lyc}(r,t)}{dr}
    =
    -4\pi r^2 n_{\rm H}^2(r,t)\alpha_B(T_{\rm II})
    -
    \langle \sigma_{\rm dust}\rangle_{\rm H}(t)
    n_{\rm H}(r,t)N_{\rm Lyc}(r,t),
    \label{eq:photon_budget_ode}
\end{equation}
where $N_{\rm Lyc}(r,t)$ is the ionizing photon rate available at radius
$r$, $\alpha_B(T_{\rm II})$ is the case-B recombination coefficient, and
$\langle \sigma_{\rm dust}\rangle_{\rm H}(t)$ is the effective dust absorption
cross section per hydrogen nucleon for Lyman-continuum photons. Within each
radial cell $n_{\rm H}$, and hence both the dust attenuation coefficient and
the recombination sink, is taken as piecewise constant, matching the
finite-volume grid, and the photon budget is swept outward from the inner
grid boundary one cell at a time.

Dust absorption and recombination compete continuously for the same photons
as they cross a cell, rather than acting in sequence, so each cell's photon
budget update is now obtained as the exact solution of the coupled linear ODE
within the cell, rather than by a single first-order (Euler) step. Writing
the cell's dust optical depth as $\tau_i = n_{{\rm H},i}\langle
\sigma_{\rm dust}\rangle_{\rm H}\,\Delta r_i$ and the recombination
requirement to keep the cell's exact spherical-shell volume $V_i$ fully
ionized as $Q_{{\rm rec},i}=\alpha_B\,n_{{\rm H},i}^2\,V_i$, the exact
per-cell update is
\begin{equation}
    N_i = N_{i-1}\,e^{-\tau_i} - Q_{{\rm rec},i}\,\frac{1-e^{-\tau_i}}{\tau_i},
    \label{eq:discrete_photon_update}
\end{equation}
where $N_{i-1}$ is the photon rate entering the cell. The discount factor
$(1-e^{-\tau_i})/\tau_i\to1$ as $\tau_i\to0$, recovering the dust-free limit
$N_i = N_{i-1}-Q_{{\rm rec},i}$.

Cells are ionized from the center outward until the available photon budget
is exhausted. If Equation~\eqref{eq:discrete_photon_update} would drive $N_i$
to a value at or below zero, that cell is instead assigned a partial, volumetric ionization
fraction that conserves photons,
\begin{equation}
    x_{\rm tr} = \min\!\left(\frac{N_{i-1}}{Q_{{\rm rec},i}},\,1\right),
\end{equation}
using the photon rate entering the cell; this boundary-cell estimate neglects
dust competition within the partial cell itself, unlike the exact treatment
of Equation~\eqref{eq:discrete_photon_update} applied to every fully-ionized
cell in the sweep. All cells exterior to this boundary are treated as
neutral. The resulting ionized fraction profile $x_i$ is then used in
Equation~\eqref{eq:two_temperature_pressure} during the subsequent
hydrodynamic update.

This procedure assumes that the ionization state adjusts instantaneously to
the evolving density field and ionizing luminosity. This approximation is
appropriate when the recombination time is short compared to the dynamical
timescale of the H~II region. It also neglects radiative-transfer effects such
as diffuse radiation, photon leakage, and shadowing by inhomogeneous gas. These
effects can change the detailed morphology and escape fraction of ionizing
photons, but the present one-dimensional treatment captures the basic
competition between declining ionizing luminosity, pressure-driven gas
expansion, and the resulting electron columns relevant for FRB dispersion
measures.

\bibliography{bib.bib}{}
\bibliographystyle{aasjournalv7}
 


\end{document}